\documentclass[conference]{IEEEtran}
\IEEEoverridecommandlockouts
\usepackage{cite}
\usepackage{amsmath,amssymb,amsfonts}
\usepackage{algorithmicx}
\usepackage[noend]{algpseudocode}
\usepackage{algorithm}
\usepackage{wasysym}
\usepackage{array}
\usepackage{url}
\usepackage{graphicx}
\usepackage{textcomp}
\usepackage{xcolor}
\usepackage{listings}
\usepackage{paralist}
\def\BibTeX{{\rm B\kern-.05em{\sc i\kern-.025em b}\kern-.08em
    T\kern-.1667em\lower.7ex\hbox{E}\kern-.125emX}}

\newcommand\lquote{\text{\textquoteleft}}
\newcommand\rquote{\text{\textquoteright}}

\begin{document}

\title{\textit{Sdft: A PDG-based Summarization for\\ Efficient Dynamic Data Flow Tracking}\thanks{to appear at QRS 2021. This is the author version.}}

\author{\IEEEauthorblockN{Xiao Kan}
\IEEEauthorblockA{
\textit{Xidian University}\\
Xi'an, China \\
814091656@qq.com
}
\and
\IEEEauthorblockN{Cong Sun$^\dagger$\thanks{$\dagger$ corresponding author}}
\IEEEauthorblockA{
\textit{Xidian University}\\
Xi'an, China \\
suncong@xidian.edu.cn}
\and
\IEEEauthorblockN{Shen Liu}
\IEEEauthorblockA{\textit{NVIDIA}\\
Santa Clara, CA, USA \\
sheliu@nvidia.com}
\and
\IEEEauthorblockN{Yongzhe Huang}
\IEEEauthorblockA{\textit{The Pennsylvania State University}\\
University Park, PA, USA \\
yzh89@psu.edu}
\and
\IEEEauthorblockN{Gang Tan}
\IEEEauthorblockA{\textit{The Pennsylvania State University}\\
University Park, PA, USA \\
gtan@psu.edu}
\and
\IEEEauthorblockN{Siqi Ma}
\IEEEauthorblockA{\textit{The University of Queensland}\\
Brisbane, Australia\\
slivia.ma@uq.edu.au}
\and
\IEEEauthorblockN{Yumei Zhang}
\IEEEauthorblockA{
\textit{Xidian University} \\
Xi'an, China \\
zhangyumei319@163.com}
}

\maketitle

\begin{abstract}
Dynamic taint analysis (DTA) has been widely used in various security-relevant scenarios that need to track the runtime information flow of programs. Dynamic binary instrumentation (DBI) is a prevalent technique in achieving effective dynamic taint tracking on commodity hardware and systems. However, the significant performance overhead incurred by dynamic taint analysis restricts its usage in production systems. Previous efforts on mitigating the performance penalty fall into two categories, parallelizing taint tracking from program execution and abstracting the tainting logic to a higher granularity. Both approaches have only met with limited success.

In this work, we propose Sdft, an efficient approach that combines the precision of DBI-based instruction-level taint tracking and the efficiency of function-level abstract taint propagation. First, we build the library function summaries automatically with reachability analysis on the program dependency graph (PDG) to specify the control- and data dependencies between the input parameters, output parameters, and global variables of the target library. Then we derive the taint rules for the target library functions and develop taint tracking for library function that is tightly integrated into the state-of-the-art DTA framework Libdft. By applying our approach to the core C library functions of glibc, we report an average of 1.58x speed up of the tracking performance compared with Libdft64. We also validate the effectiveness of the hybrid taint tracking and the ability on detecting real-world vulnerabilities.
\end{abstract}


\section{Introduction}

Dynamic taint analysis (DTA), also known as dynamic data-flow tracking (DFT), is a powerful technique for tracking information flows of software at runtime and has been used widely in vulnerability detection, program protection, information flow control, and reverse engineering.
Without accessing the source code, binary-level dynamic data-flow tracking usually uses some runtime techniques, e.g., dynamic binary instrumentation (DBI), virtual machine monitor, or whole system emulator, to monitor the target program transparently and propagate sensitive taints across memory and program contexts. Then the knowledge on taint propagation is collected directly over the target binary.

The DBI-based dynamic taint analyses \cite{DBLP:conf/ndss/NewsomeS05, DBLP:conf/iscc/ChengZYH06, DBLP:conf/micro/QinWLKZW06, DBLP:conf/issta/ClauseLO07, DBLP:conf/vee/KemerlisPJK12} are promising and flexible to track in-process tainting behaviors. Such approaches mainly focus on direct data flows and hold the tainting states within tagging memory. The key feature is to track memory locations and CPU registers that store sensitive or suspicious data. This kind of tainted data is propagated and checked at particular program execution points to decide if specific runtime properties are satisfied, e.g., whether some pointer in instruction is controlled and tampered with by attackers. To generalize the DBI-based approaches, several extensions have addressed implicit data flows \cite{DBLP:conf/ndss/KangMPS11}, flows among multiple processes \cite{DBLP:conf/IEEEares/KimKCS09}, or the generalization of taint propagation semantics to more instruction set architectures (ISA) \cite{DBLP:conf/ndss/ChuaWBSLS19}. Libdft \cite{DBLP:conf/vee/KemerlisPJK12, DBLP:conf/ndss/JeePKGAK12} and its 64-bit reimplementation \cite{libdft64} are the leading DBI-based taint tracking approach that has a relatively moderate slowdown to the native execution \cite{DBLP:conf/sac/MallisseryWHB20, DBLP:conf/ccs/GaleaK20}. This efficient dynamic taint analysis has been used to capture the data provenance \cite{DBLP:conf/ipaw/Stamatogiannakis14} or the common characteristics of valid inputs of gray-box fuzzing \cite{DBLP:conf/ndss/0001JKCGB17, DBLP:conf/acsac/Jain0GB18}.

The significant performance penalty of dynamic taint analysis has been a prominent weakness for a long time. The complex taint tracking takes a much longer time to execute the instrumented program than the original program. Improvements are on two orthogonal dimensions, i.e., \emph{parallelization} or \emph{sequential abstraction}. The data tracking can be offloaded/decoupled from the program execution to introduce more parallelization over different cluster nodes, CPUs, hosts, processes, or threads \cite{DBLP:conf/osdi/QuinnDCF16, DBLP:conf/dsn/KannanDK09, DBLP:conf/ccs/JeeKKP13, DBLP:conf/uss/MingWXW015, DBLP:conf/kbse/MingWWXL16}. On the other hand, we can aggregate the taint analysis from per-instruction tracking to a higher granularity. The tainting logic of code segments is specified at a more abstract level, e.g., at each basic block or function. At the basic block level, the tainting operations are expressed with \emph{taint flow algebra} \cite{DBLP:conf/ndss/JeePKGAK12}, optimized with compiler-based techniques, and aggregated into units that rarely interfere with the target program. \emph{Fast path} \cite{DBLP:conf/micro/QinWLKZW06, DBLP:conf/cgo/SaxenaSP08, DBLP:conf/ccs/GaleaK20} optimizes dynamic taint analysis by instrumenting a check-and-switch mechanism at specific program points where the live locations are untainted to support switching to an efficient version of code without taint tracking computations. At the function level, TaintEraser \cite{DBLP:journals/sigops/ZhuJSKW11} first proposes function-level summaries for Windows kernel APIs to improve the efficiency of dynamic taint tracking. The function summaries of this approach are specified on-demand by human effort, thus preventing the usage on a large scale to the standard libraries. Automatically inferring the library summaries for the information flows has been proposed in different application scenarios, e.g., Android libraries \cite{DBLP:conf/aplas/ZhuDD13, DBLP:conf/icse/ArztB16}. Static reachability analysis on data-flow graphs has been used to identify data propagation paths \cite{DBLP:conf/sac/MallisseryWHB20}. The derived paths have been neither used to derive function summaries nor integrated into any dynamic taint analysis. By observing that the function-level taint tracking is more abstract and efficient than the instruction-level taint tracking, the work in this paper belongs to the category of sequential abstraction.

In this paper, we present Sdft, a framework that automatically derives library function summaries and taint rules to improve the efficiency of dynamic taint analysis. Sdft analyzes the library source code to generate interprocedural program dependency graphs (PDG) for the target library functions. On the PDGs, we use reachability analysis to derive the function summaries specifying the control- and data dependencies between the input parameters, the output parameters, and the global variables of the library. Then we derive the function-level taint rules for the target library functions. Because the PDG-based analysis requires source code, to address the usability, we focus on the abstraction of standard library functions whose source code is obtainable. We apply our approach to the core functions of the standard C library, i.e. glibc. We implement a dynamic binary instrumentation tool by extending the state-of-the-art dynamic taint analysis framework Libdft \cite{DBLP:conf/vee/KemerlisPJK12, libdft64}. Our dynamic taint analysis tool can switch at runtime between the instruction-level user-code taint tracking and the function-level library function taint tracking. The contributions of this paper are summarized as follows:

\begin{enumerate}
  \item We propose a PDG-based automatic function summarization and taint rule generation for modeling the tainting behaviors of library functions invoked by the applications.
  \item We tightly integrate the function-level tainting behavior abstractions into the dynamic taint analysis by developing an extension of the state-of-the-art DTA framework Libdft.
  \item We apply our approach on glibc 2.27 and evaluate Sdft on the efficiency, tainting effects, and effectiveness of vulnerability tracking. By abstracting the core functions of the standard C library, Sdft can achieve an average 1.58x speed up on performance compared with Libdft64.
\end{enumerate}

\section{Motivating Example}\label{sec:example}

We present the motivating example in Fig.~\ref{fig:code}. In the example, a simplified \verb|memcpy| is used by another function \verb|student_cpy| to copy a struct to a global struct of \verb|student|. We assume both \verb|memcpy| and \verb|student_cpy| are library functions wrapped in a shared object \verb|libcopy.so|. To facilitate data-flow tracking, we assume the standard I/O function \verb|fgets| as data source and \verb|printf| as data sink. In the user code \verb|main.c|, the critical input from \verb|fgets| is copied by the library function \verb|student_cpy| to the global \verb|student| struct and finally printed by \verb|printf|. Consequently, a sensitive flow should be captured by our dynamic taint analysis approach. The state-of-the-art instruction-level taint analysis, e.g., Libdft \cite{DBLP:conf/vee/KemerlisPJK12, libdft64}, will go through both the user code (i.e., the binary of \verb|main|) and the library code (i.e., the binary of \verb|libcopy.so|) to conduct per-instruction instrumentation and taint propagation. In contrast, we will elaborate that our approach generates effective function summary and taint rules for the library functions (i.e., \verb|memcpy| and \verb|student_cpy|) and applies the rules in the dynamic taint analysis framework to avoid the instructions in these functions being instrumented, thus improving the overall efficiency.

\begin{figure}[!t]
\centering
\begin{lstlisting}
// libcopy.so :
typedef struct{
    char id[SIZE];
    int score;
} student;
student stu;
void *memcpy(void *dest, const void *src, size_t n) {
    char *dp = dest;
    const char *sp = src;
    while (n--)
        *dp++ = *sp++;
    return dest;
}
void student_cpy(student *src){
    memcpy(stu.id, src->id, SIZE);
    stu.score = src->score;
}
// main.c :
int main(){
    student s;
    fgets(s.id, SIZE-1, stdin);
    s.score = 95;
    student_cpy(&s);
    printf("(%s: %d)\n", stu.id, stu.score);
}
\end{lstlisting}
\vspace{-1ex}
\caption{Application of a simplified \texttt{memcpy}}\label{fig:code}
\vspace{-2ex}
\end{figure}

\section{Design of Sdft}\label{sec:design}

In this section, we describe the framework of Sdft, especially on generating the summaries of functions and their taint rules used in the dynamic taint tracking of the framework.

\subsection{System Overview}

\begin{figure}[!t]
\centering
\includegraphics[width=3.5in]{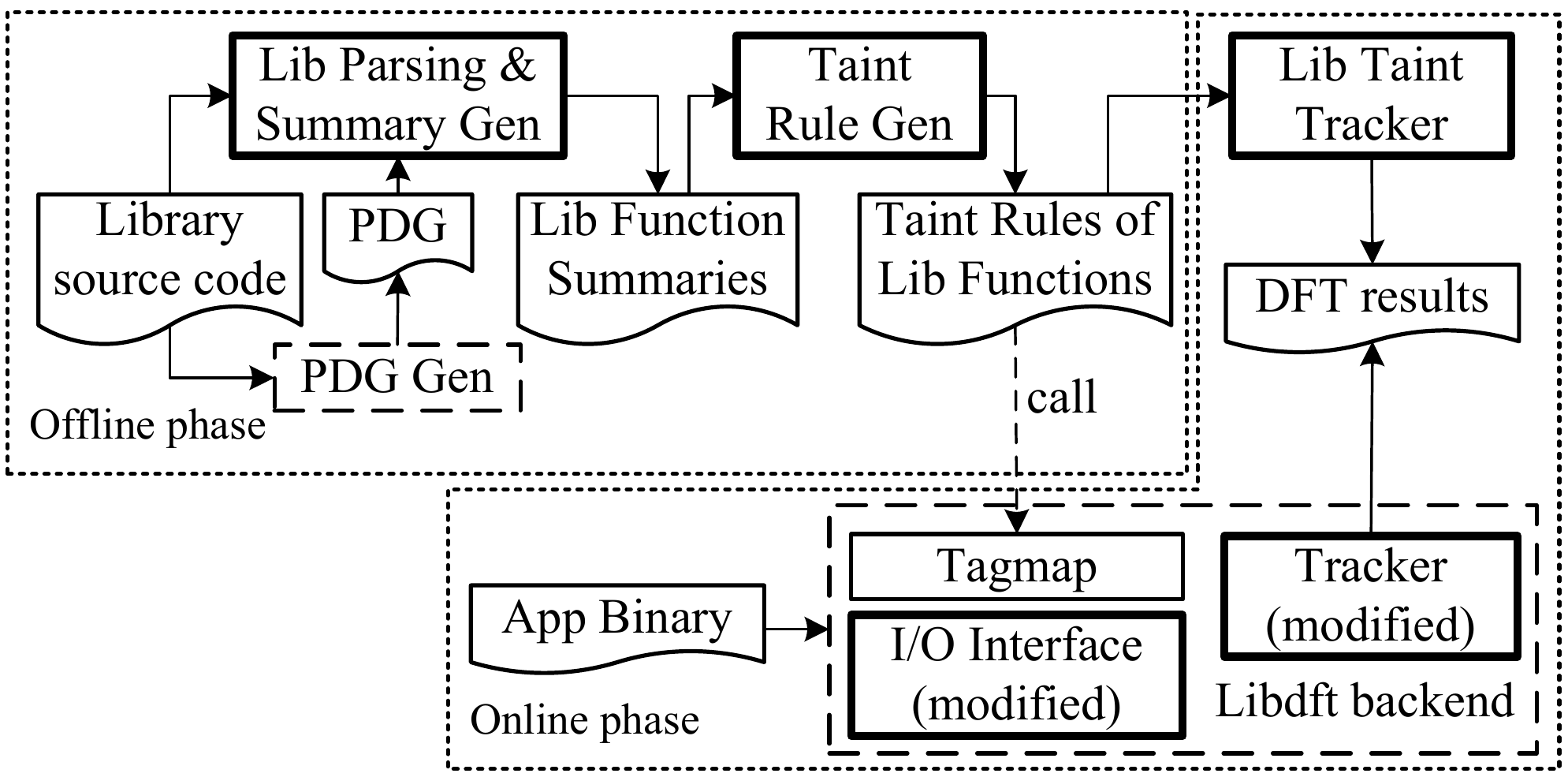}
\caption{Framework of Sdft}\label{fig:sdft-arch}
\vspace{-2ex}
\end{figure}

\emph{Key Concepts of Libdft.} As the state-of-the-art DBI-based dynamic taint analysis framework, Libdft \cite{DBLP:conf/vee/KemerlisPJK12} and its descendant Libdft64 \cite{libdft64} are Pintools developed on top of the Intel DBI framework Pin \cite{DBLP:conf/pldi/LukCMPKLWRH05}. Libdft has three main components: \textit{Tagmap}, \textit{tracker}, and \textit{I/O Interface}, as shown in Fig.~\ref{fig:sdft-arch}. The \textit{Tagmap} provides memory space and operation interface of the shadow memory and shadow registers, which models the runtime tainting state of the program. The \textit{tracker} instruments the binary program with proper analysis routines before each instruction to operate on the taint tags according to the data-flow tracking logic of each instruction. The \textit{I/O Interface} handles the taint propagation and sanitization of system calls. For each system call, this component conducts a pre/post-call-site instrumentation that inserts stubs running in user mode.

\emph{The Framework of Sdft.} As presented in Fig.~\ref{fig:sdft-arch}, the framework of Sdft consists of two phases: the \emph{offline phase} and the \emph{online phase}. The offline phase generates the taint rules from the source code of library functions. In this phase, we firstly use an off-the-shelf PDG generator \cite{DBLP:conf/ccs/LiuTJ17} to generate an interprocedural PDG for library functions. We parse the library headers in source code to flatten complex data types (Section~\ref{subsec:parsing}) and then generate the function summaries with reachability analysis (Section~\ref{subsec:summary-gen}). Then, we derive the taint rules of library functions (Section~\ref{subsec:design-taint-rule-gen}) as the critical configuration of the online data-flow tracking. On the other hand, the online phase launches the execution of binary and conducts dynamic data-flow tracking. For this phase, we propose an extension on Libdft64 to support an instruction-level and function-level interleaving data-flow tracking. The extension includes a new module of taint tracker and modifications on the main modules of Libdft64, as demonstrated in Section~\ref{sec:implementation}.

\begin{figure}[!t]
\centering
{\small
\begin{align*}
sign ::= & \texttt{signed} \mid \texttt{unsigned} & \text{(Signedness)}\\
intsz ::= & \texttt{I8} \mid \texttt{I16} \mid \texttt{I32} \mid \texttt{I64} & \text{(Integer size)} \\
floatsz ::= & \texttt{F32} \mid \texttt{F64} & \text{(Float size)} \\
\tau ::= & \texttt{int}(intsz, sign) \mid \texttt{float}(floatsz) \mid \\
& \texttt{char} \mid \texttt{void} \mid \texttt{pointer}(\tau) \mid \\
& \texttt{array}(\tau, n) \mid \texttt{function}(\tau*, \tau) \mid \\
& \texttt{struct}(id, \phi) \mid \texttt{union}(id, \phi) & \text{(Type)} \\
\phi ::= & (id, \tau)* & \text{(Field list)} \\
consDecl ::= & \texttt{struct}(id, \phi) \mid \texttt{union}(id, \phi) & \text{(Cons Decl)}\\
glbStmt ::= & \texttt{extern }\tau\ gvar; & \text{(Gvar Decl)} \\
funStmt ::= & \texttt{function}(\tau_0,\ldots, \tau_{n-1}, \tau_\text{ret})\ fid; & \text{(Func Decl)} \\
H ::= & (consDecl \mid glbStmt \mid funStmt)* & \text{(Headers)}
\end{align*}
}%
\vspace{-4ex}
\caption{Abstract syntax of C language header}\label{fig:header-ast}
\vspace{-3ex}
\end{figure}

\subsection{Library headers parsing}\label{subsec:parsing}

We take a similar abstract syntax of the C language types as in \cite{DBLP:journals/jar/BlazyL09}. The abstract syntax of types and the library headers are in Fig.~\ref{fig:header-ast}. For the type alias in the library defined as \verb|typedef|$(\tau',\tau)$, we assume the alias $\tau$ has been eliminated by static substitution. For a specific library $\ell$, the library headers $H_\ell$ consist of a set of struct or union declarations $\mathcal{C}_\ell$, a list of global variable definitions $\mathcal{G}_\ell$, and a list of API declarations $\mathcal{F}_\ell$. In our static function summary generation, unions are treated conservatively in the same way as structs. Therefore in the following, we also use \verb|cons|$(id,\phi)$ to stand for $\texttt{struct}(id, \phi)$ or $\texttt{union}(id, \phi)$. Treating union in the same way as struct will not bring in imprecision because in the well-formed library definition, even we enumerate all the members of a union as the input or output of a function, only one member will be used in the tainting procedure at runtime. The rest members will not receive or propagate taints and their taint rules will not be used. We use Algorithm~\ref{algo:flatten_types} to flatten complex types (unions and structs) to a set of primitive types. In the algorithm, $\textit{isPrimType}(\tau)$ decides if the type $\tau$ is \verb|int|, \verb|float|, \verb|char|, \verb|void|, or the pointer of these types. The primitive types of each struct or union can be retrieved from $primTypeMap$ with the $id$ of struct or union.

\begin{algorithm}[t]
\caption{Flatten Complex Types to Primitive Types}\label{algo:flatten_types}
\begin{algorithmic}[1]\small
\Procedure{FlattenPrimTypes}{$\mathcal{C}_\ell$}
    \State $types \gets \emptyset, primTypeMap \gets \emptyset$
    \ForAll{$\texttt{cons}(id,\phi)\in \mathcal{C}_\ell$}{}
        \State \Call{CollectPrimType}{$\texttt{cons}(id,\phi), types$}
        \State $primTypeMap.put(id\mapsto types)$
    \EndFor
    \State \textbf{return } $primTypeMap$
\EndProcedure

\Procedure{CollectPrimType}{$\texttt{cons}(\_,\phi), typeSet$}
    \State $\text{Suppose }\phi\equiv \{(id_1,\tau_1),\ldots,(id_k,\tau_k)\}$
    \ForAll{$\tau\in\{\tau_1,\ldots,\tau_k\}$}{}
        \If{$\textit{isPrimType}(\tau)$}
            \State $typeSet.add(\tau)$
        \ElsIf{$\tau\equiv \texttt{cons}(id',\phi')$}
            \State \Call{CollectPrimType}{$\tau, typeSet$}
        \EndIf
    \EndFor
\EndProcedure
\end{algorithmic}
\end{algorithm}

\subsection{PDG-based Summary Generation}\label{subsec:summary-gen}

The summary of a function specifies the control- and data dependencies between the input parameters, the output parameters, and the global variables of the library. Such relations are built with a PDG. The interprocedural PDG of each library function is derived using PtrSplit \cite{DBLP:conf/ccs/LiuTJ17}. In the PDG of each function, we choose proper nodes and apply reachability analysis to derive the summaries. Firstly, we define the following predicates that operate on the PDG:
\begin{enumerate}
  \item \textit{findPath}($n_{s}, n_{t}$): Depth-first traversal from node $n_s$ to $n_t$ in PDG, returning all the paths consisting of the edges with type \verb|D_gnrl|, \verb|DEF_USE|, or \verb|RAW|, as defined in Table~\ref{tab:pdg_edge_def}.
  \item \textit{findNode}($ins,p$): For the LLVM-IR instruction $ins$ in function $p$, the predicate gets the node of $ins$ in the PDG of function $p$.
  \item \textit{findNextUse}($ins$): Returning the target node of a \verb|DEF_USE| edge, if the LLVM-IR instruction $ins$ is in the source node of this edge. If $ins$ assigns a value to variable $\%v$, the predicate returns the nearest subsequent instruction in the current function that uses $\%v$.
\end{enumerate}
Intuitively, an edge $(n_i,n_j)$ in a path found by \textit{findPath} indicates $n_j$ is data-dependent to $n_i$. The path returned by \textit{findPath} specifies a relationship between the source and target node s.t. when the source node defines variable/parameter, computes a new value, or reads memory locations, how the data involved propagate, and how they are interfered with by the taint data in the execution of the function. We do not consider the data dependency edges with type \verb|D_ALIAS| since they may trigger many irrelevant circular paths. To figure out the implicit flows in some library functions as mentioned in Section~\ref{subsec:tainting_effect}, we also integrate the control dependency edge identification as an option of \textit{findPath}.

For the reachability analysis, we first decide the source nodes and target nodes for the PDG of each library function. The source and target node respectively represent the input and output of each library function. For a library function $p$, Let $N_p^s$ and $N_p^t$ be the set of source and target nodes in its PDG. To derive $N_p^s$, we first collect the candidate node of parameters/global variables. Then we add the primitive-type candidates into $N_p^s$. For the struct or union candidate, if some of its fields are used, i.e., loaded by some \verb|load| instruction under a specific instruction sequence pattern, we add the node of loading action into $N_p^s$. To derive $N_p^t$, we first label the return instruction of function $p$ as a target node in $N_p^t$. Then, if we find some primitive-type global variable used in $p$ and stored with some value, such \verb|store| instruction serves as an output of $p$. For the pointer-type global variables and parameters, if the memory pointed by these variables or parameters is modified by some \verb|store| instruction in the function under specific instruction sequence patterns, the \verb|store| instruction is added to $N_p^t$.

The summaries are defined on the parameters and return value of each function and the global variables. However, the nodes captured in $N_p^s$ and $N_p^t$ cannot always stand for the function parameter, global variable, or return instruction. We define a mapping relation $\wp$ from the source or target nodes to the function parameters, global variables, and return instructions, i.e. $\wp:N^s\cup N^t\mapsto N_{para}\cup N_{glb} \cup N_{ret}$, s.t. $N_{para}$, $N_{glb}$, and $N_{ret}$ are defined in Table~\ref{tab:pdg_node_def}. $\wp(n)$ represents the parameter, global variable, or return instruction bound to the PDG node $n$. In reality, $\wp$ is complicated to support fine-grained relations from the source/target nodes to some field of struct parameters. We take ad-hoc instruction pattern analysis to figure out these relations.

\begin{figure}[!t]
\centering
\includegraphics[width=3in]{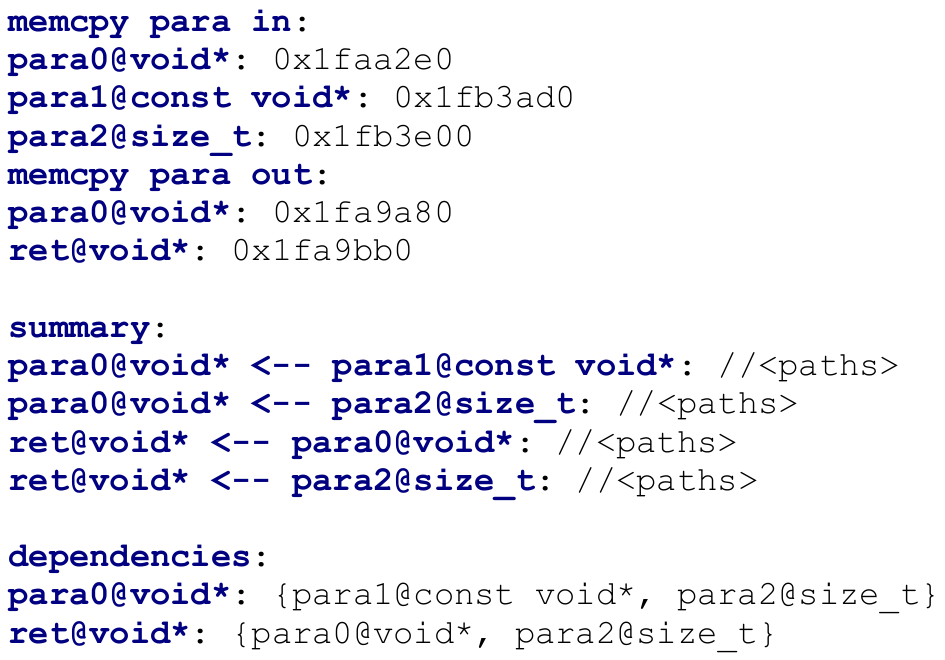}
\caption{Function summaries of \texttt{memcpy} of Fig.~\ref{fig:code}}\label{fig:summary}
\vspace{-2ex}
\end{figure}

After collecting the source and target nodes in the PDG of each library function, we compute the data- and control dependencies between the source and target nodes with reachability analysis. For each source node $n_s\in N_p^s$ and each target node $n_t\in N_p^t$, if we find a path from $n_s$ to $n_t$, we construct the summary relation over $\wp(n_s)$ and $\wp(n_t)$. For clarity to specify the summaries, we aggregate the inputs that each output depends on, i.e. $summaries ::= \{\langle n_i^{out},\{n_{i,1}^{in},\ldots, n_{i,k_i}^{in}\}\rangle_{i=1..m}\}$. For the function \verb|memcpy| in Fig.~\ref{fig:code}, the generated summaries are in Fig.~\ref{fig:summary}.  The summaries begin with the input and output parameters and global variables captured by $\wp$ on the source and target nodes. The id of the PDG node is attached to each parameter and variable. The specific paths with the same source and target node are organized together following the source-target pair and are omitted in Fig.~\ref{fig:summary} for simplicity. For the summary of \verb|student_cpy|, the procedure \textit{findPath} takes the summary of the callee function \verb|memcpy| as input to build its dependencies. This modular treatment makes the procedure of summary generation efficient compared with the global PDG traversal.

\subsection{Taint Rule Generation}\label{subsec:design-taint-rule-gen}

The back-end dynamic tainting engine usually holds shadow memory and shadow registers for tracking the tainting behavior of the program.
When a specific memory space or register is tainted, the tag value of the respective shadow memory or shadow register is set.
To specify the tainting behaviors of function for our approach, the taint rules of each function are a kind of relation between the elements of shadow memory and shadow registers.
More specifically, this relation can be modeled as $\mathcal{R}_p=(\langle\mu,\nu\rangle, \langle\mu',\nu'\rangle)$. $\mu$ and $\mu'$ are respectively the state of shadow memory before and after the execution of function $p$. $\nu$ and $\nu'$ are the state of shadow registers before and after the execution of $p$. For a library function, we map the elements of $\langle\mu,\nu\rangle$ to the function input and map the function output to the elements of $\langle\mu',\nu'\rangle$. Then we know through this function, the possible taints can impact which part of shadow memory or registers. The taint rules are derived to operate on the involved taint tags.

We assume the tagging granularity is at the byte level. For the memory region accessed by the library function, we present two predicates to facilitate the modeling of $\mathcal{R}_p$:
\begin{enumerate}
\item \textit{getTaint}($addr, sz$): For the $sz$-byte memory region at $addr$, \textit{union} the respective 1-byte tags in $\mu$ iteratively, and return the result tag byte.
\item \textit{setTaint}($addr, tag, sz$): For the $sz$-byte memory region at $addr$, set each tag byte in the respective tag region of $\mu$ with $tag$.
\end{enumerate}
Similarly, $addr$ in these primitives can be replaced by the $id$ of the general-purpose registers. Then the predicates can also work on the shadow registers in $\nu$. In these definitions, the tag granularity is in byte. Without loss of generality, our framework can also support bit-level taint tracking with a different definition of \textit{getTaint} and \textit{setTaint}.

\begin{algorithm}[t]
\caption{Taint Rules Generation of Function $p$}\label{algo:tainting}
\begin{algorithmic}[0]\footnotesize
\Procedure{TaintRuleGen}{$summaries_p$}
    \State $rules \gets \emptyset$
    \ForAll{$\langle n^{out},\{n_1^{in},\ldots, n_k^{in}\}\rangle \in summaries_p$}
         \State $tag_x\gets 0$
        \ForAll{$n_i^{in}\in\{n_1^{in},\ldots, n_k^{in}\}$}
            \If{$\textit{isPrimType}(\mathcal{T}(n_i^{in})) \wedge \mathcal{T}(n_i^{in})\notin \{\texttt{char*,void*}\}$}
                \State $rules.append($
                \State \qquad $\lquote tag_{x} \gets tag_{x}| \textit{getTaint}(\mathcal{A}(n_i^{in}),\text{sizeof}(\mathcal{T}(n_i^{in}))) \rquote)$
            \ElsIf{$\mathcal{T}(n_i^{in})\in \{\texttt{char*,void*}\}$}
                \State $rules.append($
                \State \qquad $\lquote tag_{x} \gets tag_{x}|\textit{getTaint}(\mathcal{A}(n_i^{in}),\text{strlen}(var(n_i^{in}))) \rquote)$
            \EndIf
        \EndFor
        \If{$\textit{isPrimType}(\mathcal{T}(n^{out})) \wedge \mathcal{T}(n^{out})\notin \{\texttt{char*,void*}\}$}
            \State $rules.append($
            \State \qquad $\lquote tag_o\gets \textit{getTaint}(\mathcal{A}(n^{out}),\text{sizeof}(\mathcal{T}(n^{out})))\rquote)$
            \State $rules.append($
            \State \qquad $\lquote \textit{setTaint}(\mathcal{A}(n^{out}), tag_o|tag_x, \text{sizeof}(\mathcal{T}(n^{out})))\rquote)$
        \ElsIf{$\mathcal{T}(n^{out})\in \{\texttt{char*,void*}\}$}
            \State $rules.append($
            \State $\lquote tag_o\gets \textit{getTaint}(\mathcal{A}(n^{out}),\text{strlen}((\texttt{char*})var(n^{out})) )\rquote)$
            \State $rules.append($
            \State $\lquote \textit{setTaint}(\mathcal{A}(n^{out}),tag_o|tag_x,\text{strlen}((\texttt{char*})var(n^{out})))\rquote)$
        \EndIf
    \EndFor
    \State \textbf{return }$rules$
\EndProcedure
\end{algorithmic}
\end{algorithm}

\begin{figure}[!t]
\centering
\includegraphics[width=3.5in]{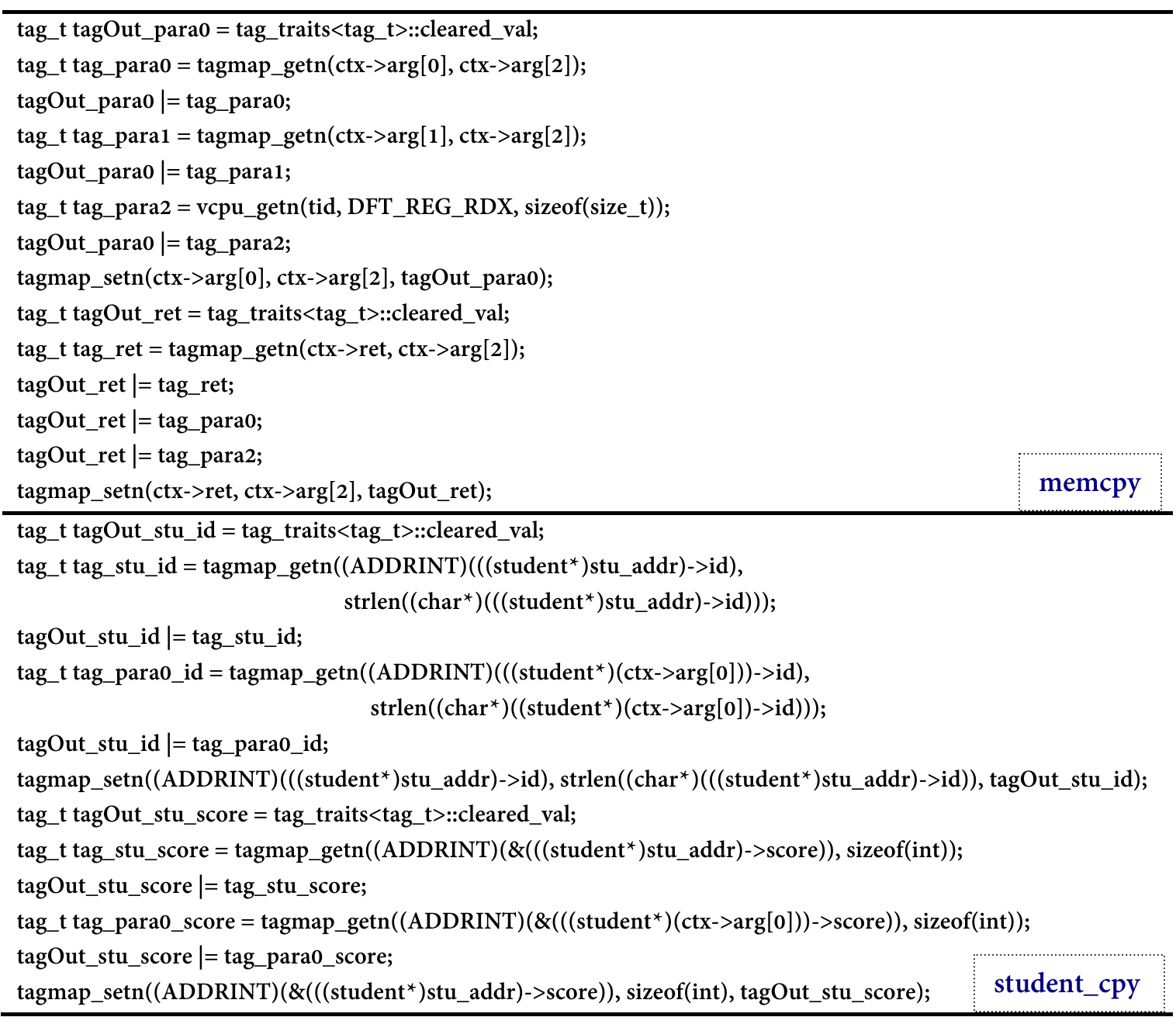}
\caption{Taint Rules of the Library Functions of Fig.~\ref{fig:code}}\label{fig:taint-rule}
\vspace{-2ex}
\end{figure}

For each node used in the summaries, it is trivial to identify the global variable name, parameter order id (argument-passing register), and return register of the node. The corresponding shadow space of such entities can be identified at runtime in our dynamic analysis. To help the online phase of Sdft identify such shadow space and conduct taint propagation, we firstly define two predicates $\mathcal{A}(n)$ and $\mathcal{T}(n)$ on PDG node $n$. For each PDG node $n\in summaries_p$, $\mathcal{A}(n)$ returns the runtime address of the global variable or parameter, or the id of register w.r.t. node $n$. $\mathcal{T}(n)$ returns the type of global variable, parameter, or return value w.r.t. node $n$. The main procedure of dynamic data-flow tracking uses the information returned by these predicates to instantiate the taint rule used for specific taint propagation action. Then, we present the taint rule generation in Algorithm~\ref{algo:tainting}. The algorithm takes the summaries as input and returns the taint rules for the specific function. The rules consist of a sequence of operations using \textit{getTaint} and \textit{setTaint} to update the tag of function output with the union of all the tags of dependent inputs. For the string or \verb|void*| array parameter, we first infer the type of element statically with the context information. If the type of element has been realized, we use \verb|strlen| to catch the bound of the tag region in the shadow memory involved in the taint propagation. Otherwise, we use a default length of array to perform the taint propagation. The default length cannot ensure the abstraction to cover the complete tag region of the parameter but can benefit the efficiency of tainting and the properly chosen length is sufficient for real-world vulnerability tracking. For the library functions in Fig.~\ref{fig:code}, the derived taint rules are listed in Fig.~\ref{fig:taint-rule}. In the rules, \textsf{tagmap\_getn} and \textsf{vcpu\_getn} respectively represent the predicate \textit{getTaint} on $\mu$ and $\nu$. \textsf{tagmap\_setn} represents the predicate \textit{setTaint} on $\mu$.

\section{Implementation Issues}\label{sec:implementation}

Sdft provides function-level tainting for library functions to improve the efficiency of the instruction-level tainting of Libdft. As shown in Fig.~\ref{fig:sdft-arch}, we develop a \textit{LibTaintTracker} component to instrument the library functions whose taint rules are generated by the offline phase (summary generation and taint rule generation). Differently from the pre/post-syscall instrumentation by \textit{I/O interface}, \textit{LibTaintTracker} performs callee-side instrumentation at the return points of the library function to insert the taint rules generated by Algorithm~\ref{algo:tainting} as a stub. Besides, \textit{LibTaintTracker} can specify a library function as a taint source or sink on demand. To the taint source function, it uses the operations of \textit{Tagmap} to introduce sensitive tags into \textit{Tagmap}. To the taint sink function, it instruments vulnerability checking code at the beginning of the function.

To implement correct switching between function-level tainting and instruction-level tainting, the instructions of the library functions should be identified to avoid being instrumented at the instruction level. To distinguish the user-code instructions from the instructions inside the library functions, we define a global flag to reserve if the current context is inside or outside library functions. We modify the module \textit{tracker} of Libdft to take the value of this flag to decide whether to instrument the instructions of the current basic block. Besides, we also modify the \textit{I/O Interface} component to avoid instrumenting the system calls invoked inside the library functions. In our implementation, \textit{I/O Interface} contains the abstraction of 335 system calls of Linux kernel v5.4.0-72.

To be more specific on generating the taint rules with Algorithm~\ref{algo:tainting}, the predicates \textit{getTaint} and \textit{setTaint} are mapped to the operation API of \textit{Tagmap}, and the taint rules are mapped to the sequences of \textit{Tagmap} operations. The implementation of predicate $\mathcal{A}()$ depends on the application binary interface (ABI) of the binaries under instrumentation. In this work, the target ISA of Sdft is x64 and the ABI we use is System V ABI. However, the taint rules generated by Algorithm~\ref{algo:tainting} are ABI-neutral. The calling convention decides how $\mathcal{A}()$ maps the PDG node to the memory locations or the registers. For example, the nodes of the first six integer parameters are mapped to the general-purpose register \verb|rdi|, \verb|rsi|, \verb|rdx|, \verb|rcx|, \verb|r8|, \verb|r9|. The nodes of SSE parameters are mapped to \verb|xmm0| to \verb|xmm7|. For the global variables, the DBI framework cannot provide sufficient addressing information for $\mathcal{A}()$. We find the offset of each global variable statically in the shared object file of the library. The address of the global variable can be derived by adding this offset to the base address of the library returned by the DBI framework with Pin's \verb|IMG_LowAddress()|.

For the variadic library functions, several arguments cannot directly map to the parameter nodes of the function's PDG. Our instrumented code snippet figures out the number of arguments and provides parameterized and iterative taint rules that take the number of arguments as a parameter.  We mainly focus on the variadic functions in \verb|stdio.h| of the C standard library. For these functions, we take heuristics to capture the number and type of arguments by analyzing the format strings in the binary. Then we iteratively get the stack address of these parameters using Pin's API.

\section{Evaluation}\label{sec:evaluation}

Our experiments are conducted on a Desktop with a 3.2GHz$\times$4 Intel Core(TM) i5-6500 CPU, 8GB RAM, Linux 5.4.0-72 kernel (Ubuntu 18.04 64-bit). The binaries are compiled with GCC v7.5.0. The PDG \cite{pdg-repo, DBLP:conf/ccs/LiuTJ17, liu2019program, liuthesis} depends on LLVM v9.0.0. The DBI framework is Pin v3.11 and Libdft64 we used is commit id 729c1b2 \cite{libdft64}.

The benchmark programs we use to evaluate the performance and effectiveness of Sdft are presented in Table~\ref{tab:benchmark}.
We compile the binaries on their default compiler optimization level. For each program, we feed the execution of the instrumented program with specific inputs to conduct the following evaluations. Specifically, for the SPEC2k6 benchmarks, we use the standard workload \verb|test| in the instrumented execution. For the other benchmarks, we enumerate the command options and launch multiple instrumented executions in a batch job using different options. For each command option, we only feed with several common arguments. This setting will not trigger a high code coverage of each benchmark but is proper to demonstrate the performance of different approaches under routine usages. We count the number of instructions in the static binaries, i.e., \#Instr in Table~\ref{tab:benchmark}. We also profile the code coverage of each benchmark under the instrumented executions, i.e.,
\begin{equation}
\frac{\text{\#BBL}_\text{exec}}{\text{\#BBL}_\text{total}}\times 100\%
\end{equation}
where $\text{\#BBL}_\text{exec}$ is the number of basic blocks in the CFG reached by the execution under the specific inputs and $\text{\#BBL}_\text{total}$ is the total number of basic blocks in the CFG of binary. The code coverage of benchmark programs ranges between 11.6\% to 64.1\%.

\begin{table}[!t]
\renewcommand{\arraystretch}{1.3}
\caption{Benchmarks and Code Coverage under Typical Inputs}
\label{tab:benchmark}
\centering
\begin{tabular}{c|r|r}
\hline
Program(version) & \#Instr & Code Coverage(\%) \\
\hline
400.perlbench(SPEC2k6) & 305,050 & 32.3 \\

401.bzip2(SPEC2k6) & 23,466 & 63.4 \\

403.gcc(SPEC2k6) & 916,519 & 31.2 \\

thttpd(2.25) & 12,499 & 17.9 \\

wget(1.19) & 66,974 & 14.5 \\

nginx(1.9.5) & 175,294 & 11.6 \\

MySQL(5.7.33) & 111,743 & 21.7 \\

FireFox(4.0) & 10,481 & 64.1 \\

\hline
\end{tabular}
\vspace{-2ex}
\end{table}

\subsection{Instantiated instrumentation}\label{subsec:instance-instrument}

Our instrumentation has been applied to a subset of library functions of glibc v2.27. The main reason that we summarize the library functions instead of the user functions is that our PDG-based analysis requires source code and the source code of library functions is far easier to be obtained compared with the source code of the user program. Indeed, the shared object file \verb|libc.so| of glibc 2.27 has 1833 library functions. Instrumenting all of these functions with Sdft will cause a very long loading time of \verb|libc.so| in the execution of the instrumented target program. More practically, we choose to instrument a major subset of the standard C library functions, which include 98 standard C library functions. All of them are declared in \verb|stdio.h|, \verb|stdlib.h|, \verb|string.h|, or \verb|time.h|.

\begin{table}[!t]
\renewcommand{\arraystretch}{1.3}
\caption{Representativeness of Instrumented Library Functions}
\label{tab:call-proportion}
\centering
\begin{tabular}{c|r|r|r}
\hline
 & \#calls & \#calls &  \\
Program & \texttt{libc.so} funcs  & candidate funcs & (\%) \\
 & ($\times 10^7$) & ($\times 10^7$) \\
\hline
400.perlbench & 2.54 & 2.22 & 87.5 \\

401.bzip2 & 1.60 & 1.44 & 90.2 \\

403.gcc & 2.36 & 2.08 & 88.3 \\

thttpd & 10.16 & 5.74 & 56.5 \\

wget & 4.02 & 1.41 & 70.1 \\

nginx & 25.62 & 23.77 & 92.7 \\

MySQL & 15.16 & 14.42 & 95.1 \\

FireFox & 0.22 & 0.20 & 90.8 \\

\hline
\end{tabular}
\vspace{-2ex}
\end{table}

\begin{table}[!t]
\renewcommand{\arraystretch}{1.3}
\caption{Number of Summaries and Taint Rules of Standard C Library Functions}
\label{tab:summary-count}
\centering
\begin{tabular}{c|c|r|r|r|r|r}
\hline
Header file & \#func & \multicolumn{4}{c|}{Avg. \#summary per func} & Avg. \#rule \\
\cline{3-6}
& & $p_{i}\rightarrow$ & $p_i\rightarrow$ & $g_i\rightarrow$ & $g_i\rightarrow$ & per func \\
& & $p_{o}$ & $g_o$ & $p_o$ & $g_o$ & \\
\hline
\texttt{stdio} & 41 & 2.00 & 0.27 & 0.17 & 0.00 & 10.90 \\

\texttt{stdlib} & 26 & 1.73 & 0.00 & 0.00 & 0.00 & 7.73 \\

\texttt{string} & 22 & 2.59 & 0.00 & 0.00 & 0.00 & 10.14 \\

\texttt{time} & 9 & 5.78 & 0.00 & 0.00 & 0.00 & 20.78 \\
\hline
Total & 98 & 2.41 & 0.11 & 0.07 & 0.00 & 10.80 \\
\hline
\end{tabular}
\vspace{-2ex}
\end{table}

We evaluate this instantiated instrumentation from the following aspects. Firstly, to demonstrate the representativeness of the 98 candidate functions, we investigate the proportion of calls to the candidate functions in the calls to all the glibc functions in the batch-job executions of the benchmark programs. Table~\ref{tab:call-proportion} presents the proportions. We can see the 98 candidate functions take {56.5$\sim$95.1\%} calls to the \verb|libc.so|. These functions have not dominated the calls to \verb|libc.so| in \textit{wget} and \textit{thttpd} because, in these programs, several advanced library functions, e.g. \verb|wcwidth| and \verb|str(n)casecmp_l|, are called frequently. These functions are not included in our function-level instrumentation but can be instrumented on-demand without much effort. The glibc functions other than these 98 candidates are instrumented per instruction in the same way as user-code functions.

Then, we present the average number of summaries and taint rules for each candidate function classified by different headers in Table~\ref{tab:summary-count}. We categorize the summaries of each function into four types: input parameters to output parameters ($p_i\rightarrow p_o$), input parameters to global variables ($p_i\rightarrow g_o$), global variables to output parameters ($g_i\rightarrow p_o$), and among global variables ($g_i\rightarrow g_o$). In Table~\ref{tab:summary-count}, we take an approximation to count the number of variadic arguments as constant. For example, for \verb|printf(const char *format, ...)|, we count the number of arguments as two to derive the function summaries. We found most of the function summaries (236 out of 254) are from input parameter to output parameter, i.e., 2.41 on average of the 98 candidate library functions. The 11 summaries of type $p_i\rightarrow g_o$ and 7 summaries of type $g_i\rightarrow p_o$ are all from \verb|stdio.h|. We have not found any dependency among global variables ($g_i\rightarrow g_o$) caused by the candidate library functions. We generate 1058 taint rules from the function summaries. After deriving the summaries of the library functions, we manually check and confirm that the dependencies specified by the summaries are consistent with the official definition of the library functions \cite{c-manual}. Such static checks are coarse-grained since we do not address the runtime size of memory regions used as arguments in different calling contexts of the library function. More fine-grained function-level tainting effect evaluations are presented in Section~\ref{subsec:tainting_effect}.

\subsection{Efficiency of Sdft}

We focus on evaluating the online phase of Sdft since the generation of the taint rules is modularized on individual functions and also offline without affecting the runtime overhead of DTA. With the instrumentation of the candidate C library functions, we investigate the efficiency of Sdft's hybrid instrumentation. We compare the execution time of benchmark programs tracked by Sdft with the execution time tracked by Libdft64. The working task of each benchmark follows the standard SPEC2k6 workload or batch-job command options as mentioned at the beginning of Section~\ref{sec:evaluation}. Because the loading procedure does not introduce taint, we skip the instrumentation of \verb|ld-linux-x86-64.so| for both tools.

\begin{table*}[!ht]
\renewcommand{\arraystretch}{1.3}
\caption{Performance Improvement of Sdft Compared with Libdft64 (Time improvement=$(T_\text{Libdft64} - T_\text{Sdft})/T_\text{Libdft64}\times 100\%$. Speed up=$T_\text{Libdft64}/T_\text{Sdft}$. $\Delta$Slowdown=$(T_\text{Libdft64} - T_\text{Sdft})/T_\text{orig}$)}
\label{tab:performance}
\centering
\begin{tabular}{c|r|r|r|r|r|r|r|r|r}
\hline
Program & \multicolumn{3}{c|}{\#Instructions Executed} & \multicolumn{3}{c|}{Execution Time(s)} & \multicolumn{3}{c}{Performance Improvement}\\
\cline{2-10}
 & \#\textit{Total}($\times 10^8$) & \#\textit{Unins.}($\times 10^8$) & (\%) & orig. & Libdft64 & Sdft & Time(\%) & Speedup & $\Delta$Slowdown\\
\hline
400.perlbench & 135.07 & 11.67 & 8.6 & 26.9 & 1402.9 & 1184.7 & 15.5 & 1.18 & 8.1\\

401.bzip2 & 84.73 & 8.65 & 10.2 & 23.7 & 830.8 & 752.4 & 9.4 & 1.10 & 3.3 \\

403.gcc & 109.76 & 10.92 & 10.0 & 6.9 & 372.7 & 304.7 & 18.2 & 1.22 & 9.9 \\

thttpd & 51.84 & 16.89 & 32.6 & 115.1 & 4048.2 & 1220.4 & 69.9 & 3.32 & 24.6 \\

wget & 103.60 & 16.45 & 15.9 & 152.2 & 5575.8 & 2910.9 & 47.8 & 1.92 & 17.5 \\

nginx & 90.59 & 20.74 & 22.9 & 2530.2 & 11194.7 & 7951.3 & 29.0 & 1.41 & 1.3\\

MySQL & 281.52 & 102.47 & 36.4 & 700.4 & 4776.9 & 3462.0 & 27.5 & 1.38 & 1.9 \\

FireFox & 154.44 & 8.22 & 5.3 & 40.1 & 938.1 & 839.1 & 10.6 & 1.12 & 2.5 \\
\hline
\end{tabular}
\vspace{-2ex}
\end{table*}

The results are presented in Table~\ref{tab:performance}. First, we profile the number of executed instructions of each benchmark program under the specific workloads, i.e., \#\textit{Total} in Table~\ref{tab:performance}. We run each benchmark program multiple times and record the average number of executed instructions. Libdft64 instruments all the instructions at runtime. In contrast, Sdft leaves a part of the executed instructions (in the 98 candidate library functions) uninstrumented. We profile the number of executed instructions inside the candidate library functions, i.e., \#\textit{Unins.} in Table~\ref{tab:performance}. These instructions are not instrumented at the instruction level. The number \#\textit{Unins.} is different from the number of call-sites in Table~\ref{tab:call-proportion}. Consequently, the proportions of these executed but uninstrumented instructions in the candidate library functions range between 5.3\% and 36.4\%. Specifically, taking \textit{400.perlbench} and \textit{FireFox} for example, the low ratios of uninstrumented instructions (i.e., 8.6\% and 5.3\% respectively) indicate that the 98 candidates library functions are in relatively rare usage by these benchmarks, even though the call sites of these library functions can dominate the call sites of all the glibc functions in these benchmarks (i.e., 87.5\% and 90.8\% respectively in Table~\ref{tab:call-proportion}).

Then, we record for each benchmark the original execution time (i.e. \emph{orig.} in Table~\ref{tab:performance}) and the execution times under the tracking of Libdft64 and Sdft respectively. We demonstrate the performance improvement of Sdft with different metrics. Specifically, Sdft achieves a 9.4\%$\sim$69.9\% time reduction and 1.10x$\sim$3.32x speedup on Libdft64. The average speedup is 1.58x. The slowdowns are reduced between 1.3x$\sim$24.6x. We find the ratios of uninstrumented instructions do not positively correlate with the performance improvement. For example, uninstrumenting 8.6\% of the executed instructions for \textit{400.perlbench} can cause a 15.5\% time reduction, while for \textit{MySQL}, the time reduction is only 27.5\% even we have uninstrumented 36.4\% of the executed instructions. This is because different types of instructions take diverse time costs to propagate the taints. The instructions in the library functions used by \textit{MySQL} are more likely in types that require less taint propagation time by Libdft64. The overall slowdowns reported in our work are not of the same magnitude as reported in \cite{DBLP:conf/vee/KemerlisPJK12} because we are working on 64-bit binaries and \cite{DBLP:conf/vee/KemerlisPJK12} reports slowdown on 32-bit benchmarks. The byte-level taint tracking on a 64-bit system introduces more complicated tainting operations on the shadow memory and shadow registers, which is time-consuming.

\subsection{Library-side Tainting Effect of Sdft}\label{subsec:tainting_effect}

Compared with the instruction-level taint tracking, function-level taint tracking of Sdft is more likely to introduce over-approximations in \textit{Tagmap}. In this section, we investigate the tainting effect of different approaches on the candidate library functions. The objective is to confirm that the function-level instrumentation of Sdft has not triggered significant over-approximation compared to the instruction-level instrumentation of Libdft64.

We craft one small user program for each of the 98 library functions to launch one-time execution of the function using randomly generated inputs. We apply parameter-sized tainting on the inputs, which means for each parameter, we taint the entire parameter but not some specific byte of the parameter. We enumerate the choices of tainting strategy. Taking the summary of \verb|memcpy| in Fig.~\ref{fig:summary} as an example, for the three input parameters \verb|para0|$\sim$\verb|para2|, we have $2^3=8$ different choices to taint the shadow memory of one, two, or three of the parameters. Under each tainting strategy, the user program runs to call the library function and we track the data flow with Sdft and Libdft64 respectively. After the call-site of the library function, we traverse the used pages of \textit{Tagmap}, calculate the tainted bytes in \textit{Tagmap}, average the number of tainted bytes of different strategies, and compare the results of Sdft and Libdft64. For space reason, we present the results of 20 library functions in Table~\ref{tab:tainting-effect} and summarize the results of the 98 library functions in the last line. The average tainted space of Sdft's function-level taint tracking is 1.02x of the space tainted by the instruction-level taint tracking of Libdft64.

Although function-level tainting tends to introduce over-approximation to taint more space, We observe that in many of the 98 standard library functions, the size of space tainted by Sdft is however slightly smaller than the space tainted by Libdft64. This is because the instruction-level tainting will propagate the taints through local variables. Such local variables are intermediate volatile tainted locations on the stack or in some general-purpose registers, e.g., \verb|rcx|, \verb|rdx|. Such locations are not tainted by our function-level instrumentation. Besides, our PDG-based summaries include the control dependencies that are omitted by the instruction-level tainting of Libdft64 based on direct flows. Therefore in several library functions, e.g., \verb|strlen| and \verb|puts|, Sdft can track the implicit flows from the argument to the return value (in \verb|rax| or \verb|xmm0|) while Libdft64 ignores them. We count 26 of the candidate library functions whose return value is control-dependent to the tainted arguments  as shown by the increase on the tainted return registers from 33 to 59  in Table~\ref{tab:tainting-effect}. In these cases, tainting more bytes indicates mitigation of under-approximation of direct flow tracking, instead of introducing more over-approximation.

\begin{table}[!t]
\renewcommand{\arraystretch}{1.3}
\caption{Tainting Effect of Sdft}
\label{tab:tainting-effect}
\centering
\begin{tabular}{c|r|r|c|c}
\hline
 & \multicolumn{2}{c|}{Avg. \#byte} & \multicolumn{2}{c}{return \texttt{rax/xmm0}}\\
Func Name & \multicolumn{2}{c|}{tainted in \textit{Tagmap}} & \multicolumn{2}{c}{tainted?} \\
\cline{2-5}
 & Libdft64 & Sdft & Libdft64 & Sdft \\
\hline
\verb|abs| & 6.0 & 4.0 & \checkmark & \checkmark \\

\verb|calloc| & 114.0 & 104.0 & $\times$ & $\times$ \\

\verb|fscanf| & 109.0 & 127.3 & $\times$ & \checkmark \\

\verb|fprintf| & 98.0 & 134.5 & $\times$ & \checkmark \\

\verb|fread| & 63.0 & 45.6 & \checkmark & \checkmark \\

\verb|free| & 56.0 & 64.0 & $\times$ & $\times$ \\

\verb|fwrite| & 101.3 & 49.8 & \checkmark & \checkmark \\

\verb|getchar| & 4.5 & 2.5 & \checkmark & \checkmark \\

\verb|gets| & 22.0 & 18.0 & $\times$ & $\times$ \\

\verb|ldiv| & 20.0 & 20.0 & \checkmark & \checkmark \\

\verb|malloc| & 70.0 & 68.0 & $\times$ & $\times$ \\

\verb|memcpy| & 14.0 & 12.3 & $\times$ & $\times$ \\

\verb|memset| & 13.5 & 12.1 & $\times$ & $\times$ \\

\verb|printf| & 25.0 & 28.0 & $\times$ & \checkmark \\

\verb|putchar| & 6.5 & 4.5 & \checkmark & \checkmark \\

\verb|puts| & 14.0 & 16.0 & $\times$ & \checkmark \\

\verb|scanf| & 16.0 & 19.3 & $\times$ & \checkmark \\

\verb|setbuf| & 80.0 & 108.0 & $\times$ & $\times$ \\

\verb|setvbuf| & 90.8 & 116.0 & $\times$ & $\times$ \\

\verb|strlen| & 13.0 & 17.0 & $\times$ & \checkmark \\
\hline

Total avg.(98 funcs) & 43.9 & 44.6 & 33\checkmark : 65$\times$ & 59\checkmark :39$\times$ \\
\hline
\end{tabular}
\vspace{-2ex}
\end{table}

To address the correctness of the taint rules generated by the algorithms, we use a validation approach. The idea is that conducting the taint rules should produce a set of noninterference results. We check if the runtime tainting results are consistent with the expected noninterference property \cite{DBLP:conf/csfw/BartheDR04}. Specifically, we modify each user program to invoke a twin-executions of the candidate library function, i.e., to call the library function twice sequentially. In each pair of executions, we randomly choose the candidate input parameter to be tainted. From the taint rules of this function, we infer the tainted (high) and untainted (low) outputs. Then, we feed the two executions with the same untainted inputs but different tainted inputs. We observe if the untainted outputs of the two executions differentiate to indicate a violation of noninterference. If the untainted outputs of the respective executions cannot be distinguished publicly under many random tainted inputs, we confirm the taint rules of the library function tend to comply with noninterference. For the example in Fig.~\ref{fig:code}, we call \verb|memcpy| twice and choose the parameter \verb|src| as tainted. In the context of \verb|student_cpy|, we know the global \verb|stu.id| is tainted but \verb|stu.score| is untainted. Using the same \verb|n| as the third parameter, we feed different random \verb|src| and \verb|src'| as the second parameter of respective executions. By observing \verb|stu.score| equals to \verb|stu'.score| many times, we know the taint rules in Fig.~\ref{fig:taint-rule} tend to enforce noninterference.

\subsection{Effectiveness of Sdft's Dynamic Data-Flow Tracking}\label{subsec:tracking}

\begin{table}[!t]\scriptsize
\renewcommand{\arraystretch}{1.3}
\caption{Effect of Sdft on Tracking Vulnerabilities of CVEs (\#instr=number of instructions instrumented. $T$=tracking time of Sdft. RCE=Remote Code Execution, S-OF=Stack Overflow, I-DIV=Integer Division-by-zero, HC=Heap Corruption)}
\label{tab:cve}
\centering

\begin{tabular}{c|c|c|r|r|r}
\hline
ID & Program & Type & \#Instr$_\text{Libdft64}$ & \#Instr$_\text{Sdft}$ & $T$(s) \\
 & & & $(\times 10^5)$ & $(\times 10^5)$ & \\
\hline
BID-6240 & wsmp3 & HC & 2.18 & 1.38 & 1.1 \\

CVE-2001-0414 & ntpd & RCE & 957.25 & 436.61 & 2.4 \\

CVE-2004-2093 & rsync & RCE & 5.37 & 3.33 & 0.3 \\

CVE-2005-1019 & Aeon & HC & 2.07 & 1.42 & 0.3 \\

CVE-2010-4221 & proftpd & S-OF & 21.93 & 13.12 & 1.5 \\

CVE-2013-2028 & nginx & S-OF & 14.89 & 10.97 & 1.3 \\

CVE-2013-4788 & glibc & S-OF & 1.80 & 1.16 & 0.4 \\

CVE-2016-9112 & openjpeg2 & I-DIV & 672.07 & 653.62 & 3.7 \\

\hline
\end{tabular}
\vspace{-3ex}
\end{table}

We validate the effectiveness of Sdft by tracking the real-world vulnerabilities of CVEs triggered by public exploits. The CVEs are taken from related works \cite{DBLP:conf/vee/KemerlisPJK12, DBLP:conf/ndss/ChuaWBSLS19, DBLP:conf/ccs/GaleaK20}, as presented in Table~\ref{tab:cve}. We only report the results of the CVEs successfully deployed in our experimental environment, especially fitting the glibc version (2.27) and the LLVM required by the PDG. Out-of-date CVEs only deployable on old systems are not addressed. The validated types of vulnerabilities include remote code execution, stack overflow, division-by-zero, and heap corruption. For each case, we develop an individual Pintool over Sdft to track the vulnerability. We retrieve the inject point of the target program including the specific syscalls, functions, or program arguments as taint sources. We treat the vulnerable functions reported in the CVEs, including the suspicious variables and parameters in these functions as taint sinks. In several cases, the taint sources are not given straightforward and we have to investigate the exploit program to identify the potential taint sources. The results demonstrate that the taint tracking of Sdft can detect the sensitive flows corresponding to all the exploits of CVEs listed in Table~\ref{tab:cve}. Meanwhile, we present in Table~\ref{tab:cve} the number of instructions instrumented by Libdft64 and Sdft during the vulnerability tracking, as well as the time cost of Sdft. On CVE-2016-9112, we did not find a significant difference in the performance of Libdft64 and Sdft because openjpeg2 rarely calls standard C library functions. Instead, it calls frequently to \verb|libopenjp2.so| and \verb|libm.so|. It needs to be emphasized that we have not developed Pintools for the use-after-free vulnerabilities thus cannot deal with the CVEs reported in Table~2 of \cite{DBLP:conf/ccs/GaleaK20}.

\section{Discussion}\label{sec:vaildity}

In this section, we discuss several limitations and threats to the validity of our approach.

\emph{a) Function pointer abstraction}. The PDG-based function summarization cannot effectively resolve the function pointer as the parameter or the global function pointer usage in the library function, because the static analysis on PDG cannot decide the control transfer targets of the indirect calls. Fine-grained binary-level CFG generations, e.g., TypeArmor\cite{DBLP:conf/sp/VeenGCPCRBHAG16} and BPA\cite{DBLP:conf/ndss/KimSZT21}, can be used to refine the PDG and support the reachability analysis inside the callees. Currently, for the library functions that use indirect calls, e.g. \verb|bsearch|, \verb|qsort| and \verb|atexit| declared in \verb|stdlib.h|, we have to return to the instruction-level instrumentation that can check if the tainted data are used in the indirect calls at runtime by instrumenting at specific instructions, e.g., \verb|jmp|, \verb|call|, or \verb|ret|.

\emph{b) Conservativeness of function-level abstraction}. The static analysis of library functions cannot decide the runtime bound of memory regions passed as the argument, therefore the taint rules iteratively apply the propagation over each element of the memory region. This policy brings in over-approximation when only some element of an array parameter is tainted. We have evaluated that the overly approximate effect in the library functions is limited and it does not affect the taint tracking of vulnerabilities (Section~\ref{subsec:tainting_effect} and~\ref{subsec:tracking}). To mitigate the over-approximation, global data-flow analysis is needed to bridge the contexts of the library function calls and the parameters to infer the range or element in the array being tainted.

\emph{c) Incompleteness of instruction support}. Because the online phase of Sdft is an extension of Libdft64, both tools take the same instruction set support. For example, at the instruction level of user code, we ignored implicit flows and register \verb|EFLAGS|. Therefore, even our function-level abstraction takes control-flow dependency as an option, the overall data-flow tracking is still limited to tracking explicit flows. However, we believe our approach can be further combined with the architectural-agnostic approach \cite{DBLP:conf/ndss/ChuaWBSLS19} which has much less limitation on the ISA supports.

\emph{d) Generability of the approach}. Sdft requires the source code of functions to build the PDGs. This limitation forces us to focus on the library functions because their source codes are obtainable. The source code is used in the offline phase. The online phase of data-flow tracking does not need source code. Therefore the requirement has no impact on the deployment of DTA. Our instantiated instrumentation gets considerable improvement by instrumenting 98 of 1833 glibc-functions. Instrumenting more library functions reduces runtime overhead but raises longer loading time of library shared objects as stated in Section~\ref{subsec:instance-instrument}. Consequently, Sdft works efficiently when complicated library functions are called intensively for a long time and the loading time becomes minor or ignorable.

\section{Related Work}\label{sec:related_work}

The system-wide dynamic taint analysis generally depends on virtual machine monitor \cite{DBLP:conf/eurosys/HoFCWH06}, system/hardware emulator \cite{DBLP:conf/asplos/SuhLZD04, DBLP:conf/eurosys/PortokalidisSB06, DBLP:conf/ccs/YinSEKK07,  ermolinskiy2010towards, DBLP:conf/raid/BosmanSB11}, or hardware-based techniques \cite{DBLP:conf/micro/VachharajaniBCROBRVA04}, or FPGA \cite{DBLP:conf/isca/DaltonKK07, DBLP:conf/dsn/KannanDK09, DBLP:conf/osdi/ZeldovichKDK08}. Dynamic binary instrumentation frameworks are widely used in tracking data flows in the address space at runtime. TaintCheck\cite{DBLP:conf/ndss/NewsomeS05} uses Valgrind \cite{DBLP:journals/entcs/NethercoteS03} and TaintTrace\cite{DBLP:conf/iscc/ChengZYH06} uses DynamoRIO\cite{DBLP:phd/ndltd/Bruening04} to instrument binary for detecting overwrite attacks and format string attacks. LIFT \cite{DBLP:conf/micro/QinWLKZW06} uses another DBI framework StarDBT \cite{DBLP:conf/cgo/BorinWWA06} on Windows, with optimizations to track taints. Dytan \cite{DBLP:conf/issta/ClauseLO07} is a dynamic taint analyzer based on the Pin framework \cite{DBLP:conf/pldi/LukCMPKLWRH05}. The framework relates taint to the \verb|EFLAGS| register and propagating taints through implicit flows based on the CFG and post-dominance information. DTA++ \cite{DBLP:conf/ndss/KangMPS11} identifies the implicit flows that potentially cause under-tainting and generates targeted taint propagation rules using CFG to resolve the under-tainting. To make the DBI-based analysis more general, there are efforts to model the data flow among multiple processes \cite{DBLP:conf/IEEEares/KimKCS09} or model the taint propagation for different target ISAs \cite{DBLP:conf/ndss/ChuaWBSLS19}.

Performance overhead is the critical issue of DBI-based dynamic taint analysis. The improvements come from two aspects. The first is to offload the data tracking from program execution to introduce parallelization over different CPUs, hosts, processes, or threads. ShadowReplica \cite{DBLP:conf/ccs/JeeKKP13} decouples the taint analysis from the execution of binary with a shadow thread that communicates with the application thread using a lock-free ring buffer structure. TaintPipe \cite{DBLP:conf/uss/MingWXW015} and StraightTaint \cite{DBLP:conf/kbse/MingWWXL16} use symbolic formulas to model taint transfer on straight-lined code segments. JetStream \cite{DBLP:conf/osdi/QuinnDCF16} tracks information flow for a sequence of time segments (epochs) with separate cores in the cluster, aggregates flow data from each epoch as a streaming computation to produce the final taint state. On the other hand, static analyses have been used to aggregate the taint propagation from per-instruction to a higher granularity. To be specific, the tainting logic of code segments becomes more abstract. TaintEraser \cite{DBLP:journals/sigops/ZhuJSKW11} first proposed to build function-level summaries for Windows kernel API functions to improve the efficiency of taint tracking. The function summaries specified on-demand by human effort cannot scale up to standard libraries. Jee et al.\cite{DBLP:conf/ndss/JeePKGAK12} segregate the program logic from the data tracking logic. The taint tracking action expressed with an algebra becomes per basic block. The optimized operations are aggregated into maximized instrumentation units to minimize interference with the target program. \emph{Fast path} \cite{DBLP:conf/micro/QinWLKZW06, DBLP:conf/cgo/SaxenaSP08, DBLP:conf/ccs/GaleaK20} is a critical concept in optimizing dynamic flow tracking. In LIFT \cite{DBLP:conf/micro/QinWLKZW06}, fast paths represent the safe data propagation of the code segment. The entry state is checked and the unsafe data are confirmed not involved in the code segment. The fast path of \cite{DBLP:conf/cgo/SaxenaSP08} takes effect when all the registers are untainted. It requires no taint computation for registers and only a write operation for clearing the memory taint for memory stores. Memory load operations are checked for the possible switch to slow path when the loaded data are tainted. TaintRabbit \cite{DBLP:conf/ccs/GaleaK20} proposes a generic taint analysis that supports more complex taint labels than traditional tags. The fast paths of TaintRabbit are generated just-in-time for the case when taint is present according to the in/out taint states of basic blocks. In summary of the sequential abstraction approaches, Most of them \cite{DBLP:conf/micro/QinWLKZW06, DBLP:conf/cgo/SaxenaSP08, DBLP:conf/ndss/JeePKGAK12, DBLP:conf/ccs/GaleaK20} do not address function-level abstraction. The function summary of \cite{DBLP:journals/sigops/ZhuJSKW11} only works on windows kernel APIs, while Sdft focuses on C libraries.
Besides, the only implementation available \cite{DBLP:conf/ccs/GaleaK20} is for 32-bit binaries while Sdft targets 64-bit binaries.

\section{Conclusion}\label{sec:conclusion}

We present Sdft, an efficient dynamic taint analysis framework that adapts function-level tainting abstraction into the state-of-the-art DBI-based instruction-level taint analysis. Sdft uses reachability analysis on modular interprocedural PDGs to derive the library function summaries for the control- and data dependencies between the inputs and outputs of the function. Then Sdft generates the taint rules for each target library function. The taint rules are then used by a taint tracker of library functions. The library taint tracker is integrated into the DTA framework Libdft64, resulting in a more efficient dynamic taint analyzer. We apply Sdft on the standard C library functions of glibc to validate the efficiency and effectiveness of our approach. Future work includes resolving the function pointer parameters and indirect calls used in the library functions, inferring precise memory bound of arguments with data-flow analysis, applying our approach to domain-specific third-party libraries, and developing more Pintools over our framework to mitigate more complicated vulnerabilities, e.g., UAFs.

\appendices

\section{Definition of PDG and Algorithms in Detail}\label{appendix}

\begin{algorithm}[!ht]
\caption{Source Nodes Derivation of Function $p$}\label{algo:sources}
\begin{algorithmic}[0]\footnotesize
\Procedure{SourceNodes}{$funStmt_p, PDG_p, \mathcal{G}_\ell$}
    \State \text{Suppose }$funStmt_p\equiv \texttt{function}(\tau_0,\ldots,\tau_{n-1},\tau_{ret})\ p$
    \State $candidates \gets \emptyset$
    \For{$i = 0..n-1$}
        \State $candidates.add(\langle\textit{findNode}(\lquote\texttt{FORMAL\_IN: }i\ \tau_i\rquote, p), \tau_i\rangle)$
    \EndFor
    \ForAll{$glbStmt\in \mathcal{G}_\ell$}
        \State \text{Suppose }$glbStmt\equiv \texttt{extern }\tau\ gvar$
        \State $candidates.add(\langle\textit{findNode}(\lquote\texttt{GLOBAL\_VALUE:@}gvar\rquote, p), \tau\rangle)$
    \EndFor
    \State $N_p^s \gets \emptyset$
    \ForAll{$\langle n_s,\tau_s\rangle \in candidates$}
        \If{\textit{isPrimType}($\tau_s$)}
            \State $N_p^s.add(n_s)$
            \State $\wp.add(n_s\mapsto n_s)$
        \ElsIf{$\tau_s\equiv\texttt{cons}(id,\_)$}
            \State $ins_x\gets \lquote\% var = \texttt{getelemptr inbounds }*\tau_s,\tau_s,\ldots \rquote$
            \State $n_x\gets \textit{findNode}(ins_x ,p)$
            \ForAll{$\tau_j\in primTypeMap.get(id)$}
                \State $ins_y\gets \lquote\texttt{load }\tau_j,\tau_j* \ \% var,\texttt{ align }\text{sizeof}(\tau_j)\rquote$
                \State $n_y\gets \textit{findNode}(ins_y,p)$
                \If{$\textit{findPath}(n_s, n_x) \neq \emptyset \wedge \textit{findNextUse}(ins_x) = ins_y$}
                    \State $N_p^s.add(n_y)$
                    \State $\wp.add(n_y\mapsto n_s)$
                \EndIf
            \EndFor
        \EndIf
    \EndFor
    \State \textbf{return }$N_p^s, \wp$
\EndProcedure
\end{algorithmic}
\end{algorithm}

\begin{algorithm}[!ht]
\caption{Target Nodes Derivation of Function $p$}\label{algo:target}
\begin{algorithmic}[0]\footnotesize
\Procedure{TargetNodes}{$funStmt_p, PDG_p, \mathcal{G}_\ell$}
    \State \text{Suppose }$funStmt_p\equiv \texttt{function}(\tau_0,\ldots,\tau_{n-1},\tau_{ret})\ p$
    \State $n_r\gets \textit{findNode}(\lquote\texttt{ret }\tau_{ret} \rquote, p)$
    \State $N_p^t\gets \{n_r\}$
    \State $\wp.add(n_r\mapsto n_r)$
    \State $candidates\gets \emptyset$
    \For{$i = 0..n-1$}
        \If{$\textit{isPointer}(\tau_i)$}
            \State $candidates.add(\langle\textit{findNode}(\lquote\texttt{FORMAL\_IN: }i\ \tau_i \rquote, p), \tau_i\rangle)$
        \EndIf
    \EndFor
    \ForAll{$glbStmt\in \mathcal{G}_\ell$}
        \State \text{Suppose }$glbStmt\equiv \texttt{extern }\tau\ gvar$
        \State $n\gets \textit{findNode}(\lquote\texttt{GLOBAL\_VALUE:@}gvar\rquote, p)$
        \If{$\textit{isPrimType}(\tau)$}
            \State $n'\gets $
            \State \qquad $\textit{findNode}(\lquote \texttt{store }\tau,\tau*@gvar, \texttt{align }\text{sizeof}(*\tau)\rquote,p)$
            \State $N_p^t.add(n')$
            \State $\wp.add(n'\mapsto n)$
        \Else
            \State $candidates.add(\langle n, \tau\rangle)$
        \EndIf
    \EndFor

    \ForAll{$\langle n_t,\tau_t\rangle \in candidates$}
        \If{\textit{isPrimPointer}($\tau_t$)}
            \State $ins_x\gets \lquote \%var=\texttt{getelemptr inbounds *}\tau_t,\tau_t, \ldots\rquote$
            \State $n_x\gets \textit{findNode}(ins_x,p)$
            \State $ins_y\gets \lquote \%var=\texttt{load }\tau_t,\tau_t*\ \_, \texttt{align }\text{sizeof}(\tau_t)\rquote$
            \State $n_y\gets \textit{findNode}(ins_y,p)$
            \State $ins_z\gets \lquote \texttt{store }\tau_t,\tau_t*\ \%var,\texttt{align }\text{sizeof}(*\tau_t)\rquote$
            \State $n_z\gets \textit{findNode}(ins_z,p)$
            \If{$(\textit{findPath}(n_t, n_x)\neq\emptyset \wedge \textit{findNextUse}(ins_x)=ins_z ) \vee$\\ \hspace{1.4cm} $(\textit{findPath}(n_t, n_y)\neq\emptyset \wedge \textit{findNextUse}(ins_y)=ins_z)$}
                \State $N_p^t.add(n_z)$
                \State $\wp.add(n_z\mapsto n_t)$
            \EndIf
        \ElsIf{$\tau_t\equiv \texttt{pointer}(\texttt{cons}(id,\_))$}
            \State $ins_x\gets \lquote \%var=\texttt{getelemptr inbounds *}\tau_t,\tau_t, \ldots\rquote$
            \State $n_x\gets \textit{findNode}(ins_x,p)$
            \ForAll{$\tau_j\in primTypeMap.get(id)$}
                \State $ins_y\gets \lquote \texttt{store }\tau_j,\tau_j*\ \%var,\texttt{align }\text{sizeof}(*\tau_j)\rquote$
                \State $n_y\gets \textit{findNode}(ins_y,p)$
                \If{$\textit{findPath}(n_t,n_x)\neq\emptyset\wedge \textit{findNextUse}(ins_x)=ins_y$}
                    \State $N_p^t.add(n_y)$
                    \State $\wp.add(n_y\mapsto n_t)$
                \EndIf
            \EndFor
        \EndIf
    \EndFor
    \State \textbf{return }$N_p^t,\wp$
\EndProcedure
\end{algorithmic}
\end{algorithm}

\begin{table*}[!t]
\renewcommand{\arraystretch}{1.3}
\caption{Type of PDG Nodes}
\label{tab:pdg_node_def}
\centering
\begin{tabular}{c|c|c}
\hline
Node Type & Specific Form (Subtype) & Description \\
\hline
$N_{para}$ & \verb|FORMAL_IN/FORMAL_OUT:| $i$ \verb|T|  & $i$th formal input/output parameter of function, with type \verb|T| \\

& \verb|ACTUAL_IN/ACTUAL_OUT:| $i$ \verb|T| & $i$th actual input/output argument of function call, with type \verb|T| \\

& \verb|T arg_pos:| $i$ $–$\verb|f_id| $j$ & $j$th field of the $i$th parameter/argument of function (call), with type \verb|T| \\
\hline

$N_{glb}$ & \verb|GLOBAL_VALUE:@| $n$ & The global variable with name $n$ \\
\hline

$N_{call}$ & \textit{ret}\verb|=call @|\textit{FuncSig} & Call site of function whose signature is \textit{FuncSig} and return value to \textit{ret} \\
\hline

$N_{ret}$ & \verb|ret T |$v$ & Return type-\verb|T| value from the function \\
\hline

$N_{ent}$ & \verb|<<ENTRY>>| \textit{func} & Entry point of function \textit{func} \\
\hline

$N_{gnrl}$ & -- & Other general LLVM IR instructions \\
\hline
\end{tabular}
\vspace{-2ex}
\end{table*}

\begin{table*}[!t]
\renewcommand{\arraystretch}{1.3}
\caption{Type of PDG Edges}
\label{tab:pdg_edge_def}
\centering
\begin{tabular}{c|c|l}
\hline
Edge Type & Subtype & Description \\
\hline

$E_{cdep}$ & -- & $(n_i,n_j)\in E_{cdep}$ iff the evaluation of $n_i$ decides whether $n_j$ can be executed \\
\hline

$E_{ddep}$ & \verb|D_gnrl| & $(n_i,n_j)\in E_{\texttt{D\_gnrl}}$ iff $n_j$ uses some value computed at $n_i$ \\

& \verb|D_ALIAS| & $(n_i,n_j)\in E_{\texttt{D\_ALIAS}}$ iff the variable of $n_i$ and $n_j$ are alias to each other \\

& \verb|DEF_USE| & $(n_i,n_j)\in E_{\texttt{DEF\_USE}}$ iff $n_j$ uses a variable that is defined in $n_i$ \\

& \verb|RAW| & $(n_i,n_j)\in E_{\texttt{RAW}}$ iff $n_j$ read some memory location written by $n_i$ \\
\hline

$E_{para}$ & \verb|P_form| &  edge binding the entry node of \textit{func} ($\in N_{ent}$) with its parameter nodes ($\in N_\texttt{FORMAL\_IN}$),  or edge from each \\
& & \verb|FORMAL_IN| node to its \verb|FORMAL_OUT| node. \verb|FORMAL_IN/FORMAL_OUT| nodes are control-dependent on the entry node \\

& \verb|P_act| & edge binding a call node of \textit{func} ($\in N_{call}$) with its argument nodes ($\in N_\texttt{ACTUAL\_IN})$, or edge from each \\
& & \verb|ACTUAL_IN| node to its \verb|ACTUAL_OUT| node. \verb|ACTUAL_IN/ACTUAL_OUT| nodes are control-dependent on the call node \\

& \verb|P_fld| & field inclusion relation from struct-typed node \verb|FORMAL_IN/FORMAL_OUT/ACTUAL_IN/ACTUAL_OUT| to the specific \\
& & field node \verb|T arg_pos:| $i$ \texttt{$-$f\_id:} $j$ \\
& \verb|P_in| & data flow from \verb|ACTUAL_IN| node to \verb|FORMAL_IN| node \\
& \verb|P_out| & data flow from \verb|FORMAL_OUT| node to \verb|ACTUAL_OUT| node \\
\hline

$E_{call}$ & -- & $(n_i,n_j)\in E_{call}$ iff the function with entry point $n_j\in N_{ent}$ is called at the call site $n_i\in N_{call}$ \\
\hline
\end{tabular}
\vspace{-2ex}
\end{table*}

The definition of PDG is mainly generalized from \cite{DBLP:conf/scam/Graf10}. For each library function $p$, the program dependency graph $G_p\equiv\langle N_p, E_p \rangle$ consists of the set of nodes $N_p=N_{para}\cup N_{glb}\cup N_{call} \cup N_{ret} \cup N_{ent} \cup N_{gnrl}$ and the set of edges $E_p=E_{cdep}\cup E_{ddep}\cup E_{para}\cup E_{call}$. The types of nodes and edges are respectively presented in Table~\ref{tab:pdg_node_def} and Table~\ref{tab:pdg_edge_def}. For the parameter with struct type, the \verb|FORMAL_IN|/\verb|FORMAL_OUT|/ \verb|ACTUAL_IN|/\verb|ACTUAL_OUT| nodes are further refined into the type node with the field nodes of struct. In such case, the edges of \verb|P_fld| defines such refinement relation. Meanwhile, the edges of \verb|P_in/P_out| apply to the relation of type nodes of struct, and the edges of \verb|P_form/P_act| apply to the relation of field nodes of struct.

We develop two procedures, i.e., $\textit{SourceNodes}(funStmt_p,$ $PDG_p, \mathcal{G}_\ell)$ and $\textit{TargetNodes}(funStmt_p, PDG_p, \mathcal{G}_\ell)$, to derive the source nodes $N_p^s$, target nodes $N_p^t$, and the mapping relation $\wp$ in Section~\ref{subsec:summary-gen}. Algorithm~\ref{algo:sources} derives the source nodes in $N_p^s$ and the related $\wp$ relation. In the algorithm, the set $candidates$ of $\langle node, type\rangle$ pairs holds each potential node of parameters/global variables with its type information. Algorithm~\ref{algo:target} identifies the target nodes $N_p^t$ representing the output of function $p$, as well as the related $\wp$ relation. The list $candidate$ holds the pointer-type global variables and parameters. \textit{isPointer}($\tau$) / \textit{isPrimPointer}($\tau$) decides if $\tau$ is a (primitive) pointer type. Algorithm~\ref{algo:summary} is the summary construction algorithm mentioned in Section~\ref{subsec:summary-gen}.

\begin{algorithm}[!t]
\caption{Summary Generation of Function $p$}\label{algo:summary}
\begin{algorithmic}[0]\footnotesize
\Procedure{SummaryGen}{$N_p^s$, $N_p^t$, $\wp$}
\State $summaries\gets\emptyset$
\ForAll{$n_s\in N_p^s$}
    \ForAll{$n_t\in N_p^t$}
        \If{$\textit{findPath}(n_s,n_t)\neq \emptyset \wedge \wp(n_s)\neq \wp(n_t)$}
            \State $summaries.add(\langle \wp(n_s),\wp(n_t) \rangle)$
        \EndIf
    \EndFor
\EndFor
\State \textbf{return }$summaries$
\EndProcedure
\end{algorithmic}
\end{algorithm}

\bibliographystyle{IEEEtran}
\bibliography{IEEEabrv,mybib}

\begin{thebibliography}{10}
\providecommand{\url}[1]{#1}
\csname url@samestyle\endcsname
\providecommand{\newblock}{\relax}
\providecommand{\bibinfo}[2]{#2}
\providecommand{\BIBentrySTDinterwordspacing}{\spaceskip=0pt\relax}
\providecommand{\BIBentryALTinterwordstretchfactor}{4}
\providecommand{\BIBentryALTinterwordspacing}{\spaceskip=\fontdimen2\font plus
\BIBentryALTinterwordstretchfactor\fontdimen3\font minus
  \fontdimen4\font\relax}
\providecommand{\BIBforeignlanguage}[2]{{%
\expandafter\ifx\csname l@#1\endcsname\relax
\typeout{** WARNING: IEEEtran.bst: No hyphenation pattern has been}%
\typeout{** loaded for the language `#1'. Using the pattern for}%
\typeout{** the default language instead.}%
\else
\language=\csname l@#1\endcsname
\fi
#2}}
\providecommand{\BIBdecl}{\relax}
\BIBdecl

\bibitem{DBLP:conf/ndss/NewsomeS05}
J.~Newsome and D.~X. Song, ``Dynamic taint analysis for automatic detection,
  analysis, and signature generation of exploits on commodity software,'' in
  \emph{{NDSS}'05}.\hskip 1em plus 0.5em minus 0.4em\relax The Internet
  Society, 2005.

\bibitem{DBLP:conf/iscc/ChengZYH06}
W.~Cheng, Q.~Zhao, B.~Yu, and S.~Hiroshige, ``{TaintTrace}: Efficient flow
  tracing with dynamic binary rewriting,'' in \emph{{ISCC}'06}.\hskip 1em plus
  0.5em minus 0.4em\relax {IEEE} Computer Society, 2006, pp. 749--754.

\bibitem{DBLP:conf/micro/QinWLKZW06}
F.~Qin, C.~Wang, Z.~Li, H.~Kim, Y.~Zhou, and Y.~Wu, ``{LIFT:} {A} low-overhead
  practical information flow tracking system for detecting security attacks,''
  in \emph{{MICRO-39}}.\hskip 1em plus 0.5em minus 0.4em\relax {IEEE} Computer
  Society, 2006, pp. 135--148.

\bibitem{DBLP:conf/issta/ClauseLO07}
J.~A. Clause, W.~Li, and A.~Orso, ``Dytan: a generic dynamic taint analysis
  framework,'' in \emph{{ISSTA}'07}.\hskip 1em plus 0.5em minus 0.4em\relax
  {ACM}, 2007, pp. 196--206.

\bibitem{DBLP:conf/vee/KemerlisPJK12}
V.~P. Kemerlis, G.~Portokalidis, K.~Jee, and A.~D. Keromytis, ``libdft:
  practical dynamic data flow tracking for commodity systems,'' in
  \emph{{VEE}'12}.\hskip 1em plus 0.5em minus 0.4em\relax {ACM}, 2012, pp.
  121--132.

\bibitem{DBLP:conf/ndss/KangMPS11}
M.~G. Kang, S.~McCamant, P.~Poosankam, and D.~Song, ``{DTA++:} dynamic taint
  analysis with targeted control-flow propagation,'' in \emph{{NDSS}'11}.\hskip
  1em plus 0.5em minus 0.4em\relax The Internet Society, 2011.

\bibitem{DBLP:conf/IEEEares/KimKCS09}
H.~C. Kim, A.~D. Keromytis, M.~Covington, and R.~Sahita, ``Capturing
  information flow with concatenated dynamic taint analysis,'' in
  \emph{{ARES}'09}.\hskip 1em plus 0.5em minus 0.4em\relax {IEEE} Computer
  Society, 2009, pp. 355--362.

\bibitem{DBLP:conf/ndss/ChuaWBSLS19}
Z.~L. Chua, Y.~Wang, T.~Baluta, P.~Saxena, Z.~Liang, and P.~Su, ``One engine to
  serve 'em all: Inferring taint rules without architectural semantics,'' in
  \emph{{NDSS}'19}.\hskip 1em plus 0.5em minus 0.4em\relax The Internet
  Society, 2019.

\bibitem{DBLP:conf/ndss/JeePKGAK12}
K.~Jee, G.~Portokalidis, V.~P. Kemerlis, S.~Ghosh, D.~I. August, and A.~D.
  Keromytis, ``A general approach for efficiently accelerating software-based
  dynamic data flow tracking on commodity hardware,'' in
  \emph{{NDSS}'12}.\hskip 1em plus 0.5em minus 0.4em\relax The Internet
  Society, 2012.

\bibitem{libdft64}
\BIBentryALTinterwordspacing
``libdft64,'' 2019. [Online]. Available:
  \url{https://github.com/AngoraFuzzer/libdft64}
\BIBentrySTDinterwordspacing

\bibitem{DBLP:conf/sac/MallisseryWHB20}
S.~Mallissery, Y.~Wu, C.~Hsieh, and C.~Bau, ``Identification of data
  propagation paths for efficient dynamic information flow tracking,'' in
  \emph{{SAC}'20}.\hskip 1em plus 0.5em minus 0.4em\relax {ACM}, 2020, pp.
  92--99.

\bibitem{DBLP:conf/ccs/GaleaK20}
J.~Galea and D.~Kroening, ``The taint rabbit: Optimizing generic taint analysis
  with dynamic fast path generation,'' in \emph{{ASIA} {CCS} '20}.\hskip 1em
  plus 0.5em minus 0.4em\relax {ACM}, 2020, pp. 622--636.

\bibitem{DBLP:conf/ipaw/Stamatogiannakis14}
M.~Stamatogiannakis, P.~Groth, and H.~Bos, ``Looking inside the black-box:
  Capturing data provenance using dynamic instrumentation,'' in \emph{5th
  International Provenance and Annotation Workshop, {IPAW}'14}, ser. Lecture
  Notes in Computer Science, vol. 8628.\hskip 1em plus 0.5em minus 0.4em\relax
  Springer, 2014, pp. 155--167.

\bibitem{DBLP:conf/ndss/0001JKCGB17}
S.~Rawat, V.~Jain, A.~Kumar, L.~Cojocar, C.~Giuffrida, and H.~Bos, ``Vuzzer:
  Application-aware evolutionary fuzzing,'' in \emph{{NDSS}'17}.\hskip 1em plus
  0.5em minus 0.4em\relax The Internet Society, 2017.

\bibitem{DBLP:conf/acsac/Jain0GB18}
V.~Jain, S.~Rawat, C.~Giuffrida, and H.~Bos, ``{TIFF:} using input type
  inference to improve fuzzing,'' in \emph{{ACSAC}'18}.\hskip 1em plus 0.5em
  minus 0.4em\relax {ACM}, 2018, pp. 505--517.

\bibitem{DBLP:conf/osdi/QuinnDCF16}
A.~Quinn, D.~Devecsery, P.~M. Chen, and J.~Flinn, ``{JetStream:} cluster-scale
  parallelization of information flow queries,'' in \emph{{OSDI}'16}.\hskip 1em
  plus 0.5em minus 0.4em\relax {USENIX} Association, 2016, pp. 451--466.

\bibitem{DBLP:conf/dsn/KannanDK09}
H.~Kannan, M.~Dalton, and C.~Kozyrakis, ``Decoupling dynamic information flow
  tracking with a dedicated coprocessor,'' in \emph{{DSN}'09}.\hskip 1em plus
  0.5em minus 0.4em\relax {IEEE} Computer Society, 2009, pp. 105--114.

\bibitem{DBLP:conf/ccs/JeeKKP13}
K.~Jee, V.~P. Kemerlis, A.~D. Keromytis, and G.~Portokalidis,
  ``{ShadowReplica:} efficient parallelization of dynamic data flow tracking,''
  in \emph{{CCS}'13}.\hskip 1em plus 0.5em minus 0.4em\relax {ACM}, 2013, pp.
  235--246.

\bibitem{DBLP:conf/uss/MingWXW015}
J.~Ming, D.~Wu, G.~Xiao, J.~Wang, and P.~Liu, ``{TaintPipe:} pipelined symbolic
  taint analysis,'' in \emph{{USENIX} Security 15}.\hskip 1em plus 0.5em minus
  0.4em\relax {USENIX} Association, 2015, pp. 65--80.

\bibitem{DBLP:conf/kbse/MingWWXL16}
J.~Ming, D.~Wu, J.~Wang, G.~Xiao, and P.~Liu, ``{StraightTaint:} decoupled
  offline symbolic taint analysis,'' in \emph{{ASE}'16}.\hskip 1em plus 0.5em
  minus 0.4em\relax {ACM}, 2016, pp. 308--319.

\bibitem{DBLP:conf/cgo/SaxenaSP08}
P.~Saxena, R.~Sekar, and V.~Puranik, ``Efficient fine-grained binary
  instrumentationwith applications to taint-tracking,'' in
  \emph{{CGO}'08}.\hskip 1em plus 0.5em minus 0.4em\relax {ACM}, 2008, pp.
  74--83.

\bibitem{DBLP:journals/sigops/ZhuJSKW11}
D.~Y. Zhu, J.~Jung, D.~Song, T.~Kohno, and D.~Wetherall, ``{TaintEraser:}
  protecting sensitive data leaks using application-level taint tracking,''
  \emph{{ACM} {SIGOPS} Oper. Syst. Rev.}, vol.~45, no.~1, pp. 142--154, 2011.

\bibitem{DBLP:conf/aplas/ZhuDD13}
H.~Zhu, T.~Dillig, and I.~Dillig, ``Automated inference of library
  specifications for source-sink property verification,'' in \emph{{APLAS}'13},
  ser. Lecture Notes in Computer Science, vol. 8301.\hskip 1em plus 0.5em minus
  0.4em\relax Springer, 2013, pp. 290--306.

\bibitem{DBLP:conf/icse/ArztB16}
S.~Arzt and E.~Bodden, ``{StubDroid:} automatic inference of precise data-flow
  summaries for the android framework,'' in \emph{{ICSE}'16}.\hskip 1em plus
  0.5em minus 0.4em\relax {ACM}, 2016, pp. 725--735.

\bibitem{DBLP:conf/pldi/LukCMPKLWRH05}
C.~Luk, R.~S. Cohn, R.~Muth, H.~Patil, A.~Klauser, P.~G. Lowney, S.~Wallace,
  V.~J. Reddi, and K.~M. Hazelwood, ``Pin: building customized program analysis
  tools with dynamic instrumentation,'' in \emph{{PLDI}'05}.\hskip 1em plus
  0.5em minus 0.4em\relax {ACM}, 2005, pp. 190--200.

\bibitem{DBLP:conf/ccs/LiuTJ17}
S.~Liu, G.~Tan, and T.~Jaeger, ``{PtrSplit}: Supporting general pointers in
  automatic program partitioning,'' in \emph{{CCS}'17}, 2017, pp. 2359--2371.

\bibitem{DBLP:journals/jar/BlazyL09}
S.~Blazy and X.~Leroy, ``Mechanized semantics for the clight subset of the {C}
  language,'' \emph{J. Autom. Reason.}, vol.~43, no.~3, pp. 263--288, 2009.

\bibitem{pdg-repo}
\BIBentryALTinterwordspacing
``{Parameter-tree based Program Dependence Graph (PDG)},'' 2017. [Online].
  Available:
  \url{https://bitbucket.org/psu_soslab/program-dependence-graph-in-llvm}
\BIBentrySTDinterwordspacing

\bibitem{liu2019program}
S.~Liu, D.~Zeng, Y.~Huang, F.~Capobianco, S.~McCamant, T.~Jaeger, and G.~Tan,
  ``Program-mandering: Quantitative privilege separation,'' in \emph{{CCS}'19},
  2019, pp. 1023--1040.

\bibitem{liuthesis}
S.~Liu, ``Quantitative privilege separation with pointer supports,'' Ph.D.
  dissertation, Pennsylvania State University, 2020.

\bibitem{c-manual}
S.~Loosemore, R.~M. Stallman, A.~Oram, and R.~McGrath, ``{The GNU C library
  reference manual},'' 1993.

\bibitem{DBLP:conf/csfw/BartheDR04}
G.~Barthe, P.~R. D'Argenio, and T.~Rezk, ``Secure information flow by
  self-composition,'' in \emph{{CSFW}'04}.\hskip 1em plus 0.5em minus
  0.4em\relax {IEEE} Computer Society, 2004, pp. 100--114.

\bibitem{DBLP:conf/sp/VeenGCPCRBHAG16}
V.~van~der Veen, E.~G{\"{o}}ktas, M.~Contag, A.~Pawlowski, X.~Chen, S.~Rawat,
  H.~Bos, T.~Holz, E.~Athanasopoulos, and C.~Giuffrida, ``A tough call:
  Mitigating advanced code-reuse attacks at the binary level,'' in \emph{{IEEE}
  {SP}'16}.\hskip 1em plus 0.5em minus 0.4em\relax {IEEE} Computer Society,
  2016, pp. 934--953.

\bibitem{DBLP:conf/ndss/KimSZT21}
S.~H. Kim, C.~Sun, D.~Zeng, and G.~Tan, ``Refining indirect call targets at the
  binary level,'' in \emph{{NDSS}'21}.\hskip 1em plus 0.5em minus 0.4em\relax
  The Internet Society, 2021.

\bibitem{DBLP:conf/eurosys/HoFCWH06}
A.~Ho, M.~A. Fetterman, C.~Clark, A.~Warfield, and S.~Hand, ``Practical
  taint-based protection using demand emulation,'' in
  \emph{{EuroSys}'06}.\hskip 1em plus 0.5em minus 0.4em\relax {ACM}, 2006, pp.
  29--41.

\bibitem{DBLP:conf/asplos/SuhLZD04}
G.~E. Suh, J.~W. Lee, D.~Zhang, and S.~Devadas, ``Secure program execution via
  dynamic information flow tracking,'' in \emph{{ASPLOS}'04}.\hskip 1em plus
  0.5em minus 0.4em\relax {ACM}, 2004, pp. 85--96.

\bibitem{DBLP:conf/eurosys/PortokalidisSB06}
G.~Portokalidis, A.~Slowinska, and H.~Bos, ``Argos: an emulator for
  fingerprinting zero-day attacks for advertised honeypots with automatic
  signature generation,'' in \emph{{EuroSys}'06}.\hskip 1em plus 0.5em minus
  0.4em\relax {ACM}, 2006, pp. 15--27.

\bibitem{DBLP:conf/ccs/YinSEKK07}
H.~Yin, D.~X. Song, M.~Egele, C.~Kruegel, and E.~Kirda, ``Panorama: capturing
  system-wide information flow for malware detection and analysis,'' in
  \emph{{CCS}'07}.\hskip 1em plus 0.5em minus 0.4em\relax {ACM}, 2007, pp.
  116--127.

\bibitem{ermolinskiy2010towards}
A.~Ermolinskiy, S.~Katti, S.~Shenker, L.~Fowler, and M.~McCauley, ``Towards
  practical taint tracking,'' UCB/EECS-2010-92, UC Berkeley, Tech. Rep., 2010.

\bibitem{DBLP:conf/raid/BosmanSB11}
E.~Bosman, A.~Slowinska, and H.~Bos, ``Minemu: The world's fastest taint
  tracker,'' in \emph{{RAID}'11}, ser. Lecture Notes in Computer Science, vol.
  6961.\hskip 1em plus 0.5em minus 0.4em\relax Springer, 2011, pp. 1--20.

\bibitem{DBLP:conf/micro/VachharajaniBCROBRVA04}
N.~Vachharajani, M.~J. Bridges, J.~Chang, R.~Rangan, G.~Ottoni, J.~A. Blome,
  G.~A. Reis, M.~Vachharajani, and D.~I. August, ``{RIFLE:} an architectural
  framework for user-centric information-flow security,'' in
  \emph{{MICRO-37}}.\hskip 1em plus 0.5em minus 0.4em\relax {IEEE} Computer
  Society, 2004, pp. 243--254.

\bibitem{DBLP:conf/isca/DaltonKK07}
M.~Dalton, H.~Kannan, and C.~Kozyrakis, ``Raksha: a flexible information flow
  architecture for software security,'' in \emph{{ISCA}'07}.\hskip 1em plus
  0.5em minus 0.4em\relax {ACM}, 2007, pp. 482--493.

\bibitem{DBLP:conf/osdi/ZeldovichKDK08}
N.~Zeldovich, H.~Kannan, M.~Dalton, and C.~Kozyrakis, ``Hardware enforcement of
  application security policies using tagged memory,'' in
  \emph{{OSDI}'08}.\hskip 1em plus 0.5em minus 0.4em\relax {USENIX}
  Association, 2008, pp. 225--240.

\bibitem{DBLP:journals/entcs/NethercoteS03}
N.~Nethercote and J.~Seward, ``Valgrind: {A} program supervision framework,''
  \emph{Electron. Notes Theor. Comput. Sci.}, vol.~89, no.~2, pp. 44--66, 2003.

\bibitem{DBLP:phd/ndltd/Bruening04}
D.~Bruening, ``Efficient, transparent, and comprehensive runtime code
  manipulation,'' Ph.D. dissertation, Massachusetts Institute of Technology,
  2004.

\bibitem{DBLP:conf/cgo/BorinWWA06}
E.~Borin, C.~Wang, Y.~Wu, and G.~Araujo, ``Software-based transparent and
  comprehensive control-flow error detection,'' in \emph{{CGO}'06}.\hskip 1em
  plus 0.5em minus 0.4em\relax {IEEE} Computer Society, 2006, pp. 333--345.

\bibitem{DBLP:conf/scam/Graf10}
J.~Graf, ``Speeding up context-, object- and field-sensitive {SDG}
  generation,'' in \emph{{SCAM}'10}.\hskip 1em plus 0.5em minus 0.4em\relax
  {IEEE} Computer Society, 2010, pp. 105--114.

\end{thebibliography}

\end{document}